\documentclass{aa}
\usepackage{graphicx}
\usepackage{amsmath}
\usepackage[]{natbib}
\bibpunct{(}{)}{;}{a}{}{,}
\DeclareTextSymbol{\degre}{OT1}{23}
\def \darwin {D{\sc arwin}}
\def \darwinsim {\textit{DarwinSIM}}

\usepackage{txfonts}

\begin{document}
   \title{Nulling interferometry: impact of exozodiacal clouds on the performance of future life-finding space missions}
   \titlerunning{Impact of exozodiacal clouds on the performance of future life-finding space missions}

   \author{D. Defr\`ere\inst{1}
          \and
          O. Absil\inst{1}
          \and
          R. den Hartog\inst{2}
          \and
          C. Hanot\inst{1}
          \and
          C. Stark\inst{3}
          }


   \institute{Institut d'Astrophysique et de G\'eophysique, Universit\'e de Li\`ege, 17 All\'ee du Six
 Ao\^ut, B-4000 Li\`ege, Belgium\\
              \email{defrere@astro.ulg.ac.be}
         \and
             Netherlands Institute for Space Research, SRON, Sorbonnelaan 2, 3584 CA, Utrecht, The Netherlands
         \and
             Department of Physics, University of Maryland, Box 197, 082 Regents Drive, College Park, MD 20742-4111, USA
             }

   \date{Received 23 July 2009; accepted 05 October 2009}

  \abstract
   {Earth-sized planets around nearby stars are being detected for the first time
    by ground-based radial velocity and space-based transit surveys.
    This milestone is opening the path toward the definition of missions able to
   directly detect the light from these planets, with the identification of
   bio-signatures as one of the main objectives. In that respect, both the
   European Space Agency (ESA) and the National Aeronautics and Space
   Administration (NASA) have identified nulling interferometry as one of
   the most promising techniques. The ability to study distant planets will
   however depend on the amount of exozodiacal dust in the habitable zone of the target stars.}
   {We assess the impact of exozodiacal clouds on the performance of an infrared nulling
    interferometer in the Emma X-array configuration. The first part of the study is
    dedicated to the effect of the disc brightness
    on the number of targets that can be surveyed and studied by spectroscopy
    during the mission lifetime. In the second part, we address the impact of
    asymmetric structures in the discs such as clumps and offset which
    can potentially mimic the planetary signal.}
   {We use the \darwinsim{} software which was designed and
    validated to study the performance of space-based nulling interferometers.
    The software has been adapted to handle images of exozodiacal discs and
    to compute the corresponding demodulated signal.}
   {For the nominal mission architecture with 2-m aperture telescopes, centrally symmetric exozodiacal
    dust discs about 100 times denser than the solar zodiacal cloud can be tolerated in order to survey
    at least 150 targets during the mission lifetime. Considering modeled resonant structures
    created by an Earth-like planet orbiting at 1\,AU around a Sun-like star, we show that this
    tolerable dust density goes down to about 15 times the solar zodiacal density for face-on
    systems and decreases with the disc inclination.}
   {Whereas the disc brightness only affects the integration time, the presence of clumps or offset are more
    problematic and can seriously hamper the planet detection. The upper limits on the tolerable exozodiacal
    dust density derived in this paper must be considered as rather pessimistic, but still give a realistic
    estimation of the typical sensitivity that we will need to reach on exozodiacal discs in order to prepare
    the scientific programme of future Earth-like planet characterisation missions.}



   \maketitle
%

\section{Introduction}
\label{sec:intro}

The possibility of identifying habitable worlds and even biosignatures from extrasolar
planets currently contributes to the growing interest about their nature and properties.
Since the first planet discovered around another solar-type star in 1995 \citep{Mayor:1995}, nearly
370 extrasolar planets have been detected and many more are
expected to be unveiled by ongoing or future search programmes.
Most extrasolar planets detected so far have been identified from
the ground by indirect techniques, which rely on observable
effects induced by the planet on its parent star. From the ground, radial velocity
measurements are currently limited to the detection of planets
about 2 times as massive as the Earth in orbits around Sun-like
and low-mass stars \citep{Mayor:2009b} while the transit method is
limited to Neptune-sized planets \citep{Gillon:2007}. Thanks to the
very high precision photometry enabled by the stable space
environment, the first space-based dedicated missions (namely CoRoT
and Kepler) are now expected to reveal Earth-sized extrasolar
planets by transit measurements as well. Launched in 2006, CoRoT (Convection Rotation and
planetary Transits) has detected its first extrasolar planets
\citep[e.g.,][]{Barge:2008,Alonso:2008,Leger:2009} and is expected to unveil about 100
transiting planets down to a size of 2\,R$_{\oplus}$ around G0V
stars and 1.1\,R$_{\oplus}$ around M0V stars over its entire
lifetime for short orbital periods \citep{Moutou:2005}.
Launched in 2009, Kepler will extend the survey to
Earth-sized planets located in the habitable zone of about 10$^5$
main sequence stars \citep{Borucki:2007}. After 4 years, Kepler
should have discovered several hundred of terrestrial planets with
periods between one day and 400 days. After this initial
reconnaissance by CoRot and Kepler, the Space Interferometry
Mission (SIM PlanetQuest) might provide unambiguously the mass of
Earth-sized extrasolar planets orbiting in the habitable zone of
nearby stars by precise astrometric measurements. With CoRoT and Kepler,
we will have a large census of
Earth-sized extrasolar planets and their occurrence rate as a
function of various stellar properties. However, even though the
composition of the upper atmosphere of transiting extrasolar
planets can be probed in favorable cases \citep[e.g.,][]{Richardson:2007},
none of these missions will directly detect the photons emitted by
the planets which are required to study the planet atmospheres and
eventually reveal the signature of biological activity.

Detecting the light from an Earth-like extrasolar planet is very challenging
due to the high contrast ($\sim$10$^7$ in the mid-IR,
$\sim$10$^{10}$ in the visible) and the small angular separation
($\sim$ 0.5 $\mu$rad for an Earth-Sun system located at 10\,pc)
between the planet and its host star. A technique that has been
proposed to overcome these difficulties is nulling
interferometry \citep{Bracewell:1978}. The basic principle is to combine the beams coming from two
telescopes in phase opposition so that a dark fringe appears on
the line of sight, which strongly reduces the stellar emission.
Considering the two-telescope interferometer initially proposed by
Bracewell, the response on the plane of the sky is a series of
sinusoidal fringes, with angular spacing of $\lambda/b$. By
adjusting the baseline length ($b$) and orientation, the transmission of the off-axis
planetary companion can then be maximised. However, even when the
stellar emission is sufficiently reduced, it is generally not
possible to detect Earth-like planets with a static array
configuration, because their emission is dominated by the thermal contribution
of warm dust in our solar system as well as around the target stars (exozodiacal cloud).
This is the reason why Bracewell proposed to rotate the
interferometer so that the planetary signal is modulated
by alternatively crossing high and low transmission regions, while
the stellar signal and the background emission remain constant.
The planetary signal can then be retrieved by synchronous
demodulation. However, a rapid rotation of the array would be difficult to
implement and as a result the detection is highly
vulnerable to low frequency drifts in the stray light, thermal
emission, and detector gain. A number of interferometer
configurations with more than two collectors have then been proposed to
perform faster modulation and overcome this problem by using phase
chopping \citep{Angel:1997,Mennesson:1997,Absil:2001}.  The
principle of phase chopping is to synthetize two different
transmission maps with the same telescope array, by applying
different phase shifts in the beam combination process. By differencing two
different transmission maps, it is
possible to isolate the planetary signal from the contributions of the
star, local zodiacal cloud, exozodiacal cloud, stray light,
thermal, or detector gain. Phase
chopping can be implemented in various ways (e.g.\ inherent and
internal modulation, \citealt{Absil:thesis}), and are now an essential
part of future space-based life-finding nulling interferometry
missions such as ESA's \darwin{} \citep{Fridlund:2006} and NASA's
Terrestrial Planet Finder \citep[TPF,][]{Lawson:2008}. The purpose of this
paper is to assess the impact of exozodiacal dust discs on the performance
of these missions. After describing the nominal performance of \darwin/TPF, the first
part of the study is dedicated to centrally symmetric
exozodiacal discs which are suppressed by phase chopping and only contribute through their shot noise.
If they are too bright, exozodiacal discs can drive the integration time
and we investigate the corresponding impact on the number of planets that can be surveyed during the mission lifetime.
In the second part, we address the impact of asymmetric structures in the discs (such as clumps and offset)
which are not canceled by phase chopping and can seriously hamper the planet detection process.


\section{D{\small ARWIN}/TPF overview}

\begin{figure}[!t]
\begin{center}
\includegraphics[angle=0, width=8 cm]{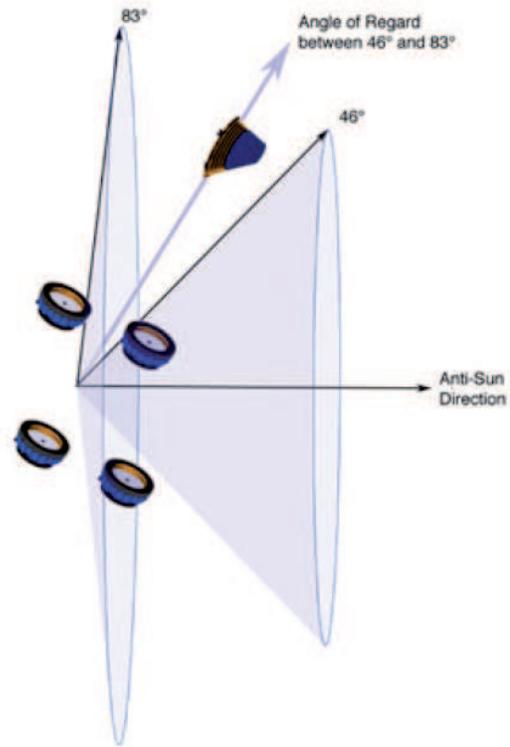}
\caption{Representation of the \darwin/TPF space interferometer in
its baseline ``Emma X-array'' configuration \citep{Leger:2007}. It
includes 4 telescopes and a beam combiner spacecraft, deployed and
observing at the Sun-Earth Lagrange point L2. At any given time,
it can observe an annular region on the sky between 46$^\circ$ and
83$^\circ$ from the anti-solar direction. During one Earth year,
this annulus rotates and gives access to almost
all regions of the celestial sphere.}\label{Fig:emmafov}
\end{center}
\end{figure}

Considerable effort have been expended in the past decade by both
ESA and NASA to design a mission that provides the required
scientific performance while minimizing cost and technical risks.
After the investigation of several interferometer architectures,
these efforts culminated in 2005-2006 with two parallel assessment
studies of the \darwin{} mission, carried out by EADS Astrium and
Alcatel-Alenia Space. Two array architectures have been thoroughly
investigated during these industrial studies: the four-telescope
X-array and the Three-Telescope Nuller \citep[TTN][]{Karlsson:2004}.
These studies included the launch requirements, payload
spacecraft, and the ground segment during which the actual mission
science would be executed. Almost simultaneously, NASA/JPL
initiated a similar study for the Terrestrial Planet Finder
Interferometer (TPF-I, \citealt{Lawson:2008}). These efforts on both sides of the
Atlantic have finally resulted in a convergence and consensus on
mission architecture, the so-called non-coplanar or Emma-type
X-array (represented in Fig.~\ref{Fig:emmafov}).

\begin{table}[t]
  \centering
  \caption{Instrumental parameters considered in this study for \darwin/TPF.}\label{Tab:darwin_spec}
  \begin{tabular}{c}
    \hline
    \hline
    \begin{tabular}{l c }
      Instrumental parameters & Emma X-Array Design\\
      \hline
      Max. baselines [m] & 400$~\times~$67 \\
      Telescope diameter [m] & 2.0 \\
      Field of regard & 46$^\circ$ to 83$^\circ$  \\
      Optics temperature [K] & 40 \\
      Detector temperature [K] & 8 \\
      Quantum efficiency & 70\% \\
      Instrument throughput$^a$ & 10\% \\
      Science waveband [$\mu$m] & 6.0-20.0 \\
      Modal filtering [$\mu$m]$^b$ & 6.0-11.5/11.5-20.0\\
      Spectral resolution & 60 \\
      Instrumental stability$^c$ & \\
      \hspace{0.4cm}- rms OPD error [nm] & 1.5 \\
      \hspace{0.4cm}- rms amplitude error & 0.05\%\\
    \end{tabular}\\
    \hline
  \end{tabular}\\
  \begin{flushleft}
    \vspace{-0.3cm}
    {\hspace{1cm}\tiny $^a$ excluding ideal beam combiner losses and coupling efficiency.}\\
    {\hspace{1cm}\tiny $^b$ see section~\ref{sec:coupling}.}\\
    {\hspace{1cm}\tiny $^c$ see section~\ref{sec:instab}.}
  \end{flushleft}
\end{table}

\subsection{Instrumental concept}

\begin{figure}[t]
\begin{center}
\includegraphics[width=8cm]{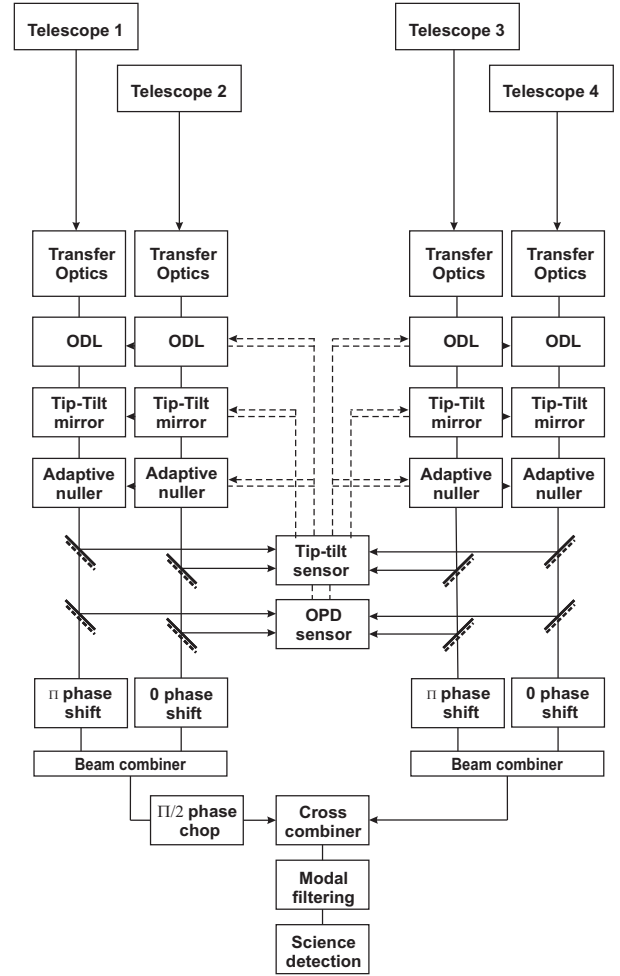}
\caption{Block diagram of the \darwin/TPF optical layout. Feed-back
signals driving the tip-tilt/OPD control are represented by dashed
lines.} \label{fig:block_diag}
\end{center}
\end{figure}

The baseline
design consists of four 2-m aperture collector spacecraft, flying
in rectangular formation and feeding light to the beam combiner
spacecraft located approximately 1200\,m above the array. This
arrangement makes available baselines up to 70\,m for nulling
measurements and up to 400\,m for the general astrophysics
programme (constructive imaging). Note that the size of the
collecting apertures has not yet been fixed and will influence
the final cost of the mission. The optical layout separates the nulling and imaging functions, the shortest baselines being used
for nulling and the longest ones for imaging. This configuration
has the advantage of allowing optimal tuning of the shorter dimension
of the array for starlight suppression while keeping a
significantly longer dimension to provide a rapid modulation of the planet
signal as the array rotates. The X-array design is also appropriate to implement
various techniques for removing instability noise, which is
one of the dominant noise contributor (see appendix~\ref{app:instab}). The assessment studies
settled on an imaging to nulling baseline ratio of 3:1, based on
scientific and instrument design constraints. A larger
ratio of 6:1 could nonetheless improve performance by simplifying noise
reduction in the post-processing of science images \citep{Lay:2006}.
The optical system architecture is represented by the block diagram in Fig.~\ref{fig:block_diag}
with the following elements in the optical path:
\begin{itemize}
\item four spherical primary mirrors located on a virtual paraboloid and focusing the beam at the paraboloid's focal point.
The virtual parabola focal length has been set to 1200\,m, resulting from a trade-off between
differential polarisation effects and inter spacecraft metrology capability, as well as to enable the
implementation of longer baselines for an imaging mode;
\item transfer optics consist of all the equipment needed to collect and redirect the incoming beams towards fixed directions whatever the interferometer configuration. This includes mainly tip-tilt mirrors to handle array reconfiguration, a derotator to handle the array rotation and a common two or three mirror telescope to achieve beam collimation;
\item optical delay lines (ODL) to adjust the optical path differences (OPD);
\item fast steering mirrors to correct for tip-tilt errors;
\item deformable mirrors to compensate for quasi-static errors such as defocus and astigmatism which are due to the use of spherical mirrors;
\item achromatic $\pi$ phase shifters to produce the destructive interference of on-axis stellar light (using Fresnel Rhombs for instance, \citealt{Mawet:2007});
\item dichroic beam splitters to separate the signals between the science waveband and the waveband used for metrology;
\item a Modified Mach Zehnder \citep[MMZ,][]{Serabyn:2001} beam combiner;
\item coupling optics to focus the outputs of the cross-combiner into single mode fibres. Chalcogenide fibres can cover successfully the
wavelength range 6.0-12 $\mu$m \citep{Ksendzov:2007} and silver halide fibres
can be used for modal filtering in at least the
10.5-17.5 $\mu$m spectral range (further investigations are however necessary to demonstrate that they are usable in the 17.5-20.0 $\mu$m wavelength range, \citealt{Ksendzov:2008});
\item a detection assembly controlled at a temperature of 8\,K and connected to the fibres.
\end{itemize}

\begin{table*}[!t]
  \centering
  \caption{Parameters adopted for the performance simulations.}
    \label{Tab:assum}
  \begin{tabular}{l l}
    \hline
    \hline
    Parameter &  \\
    \hline
    Configuration & Emma X-Array Design with a 6:1 aspect ratio (see Table~\ref{Tab:darwin_spec})\\
    Planet & Earth-sized planet with a constant temperature of 265\,K across the habitable zone  \\
    Exozodiacal density & Three times our solar system, based on Kelsall model \citep{Kelsall:1998} \\
    Mission duration & 5 years (2 years for survey and 3 years for characterisation)\\
    Retargeting time & 6 hours \\
    Integration efficiency & 70\% (accounting for overhead loss)\\
    Target stars & Darwin All Sky Survey Catalogue (DASCC, \citealt{Kaltenegger:2008})\\
    Earth-like planet per star ($\eta_\oplus$) & 1 \\
    Habitable zone & 0.7 - 1.5\,AU scaled with L$^{1/2}_\star$ \\
    Time allocation & 10\% F, 50\% G, 30\% K and 10\% M stars\\
    Array rotation period & 50000 seconds (not scaled with the array baseline length) \\
    SNR threshold for detection & 5 \\
    SNR threshold for spectroscopy & 5 (CO$_2$,O$_3$) and 10 (CO$_2$,O$_3$,H$_2$O) \\
    \hline
  \end{tabular}
\end{table*}

\darwin/TPF would be placed around the second Lagrange
point (L2) by an Ariane 5 ECA vehicle. L2 is optimal to achieve
passive cooling of the collector and beam combiner
spacecraft down to 40\,K  by means of sunshades. An additional refrigerator
within the beam combiner spacecraft cools the detector assembly to
8\,K. Due to the configuration of the array and the need for
solar avoidance, the instantaneous sky access is limited to an
annulus with inner and outer half-angles of 46$^\circ$ and
83$^\circ$ centred on the anti-sun vector (see
Fig.~\ref{Fig:emmafov}, \citealt{Carle:2005}). This annulus transits
the entire ecliptic circle during one year, giving access to
almost the entire sky.

\subsection{Scientific objectives}

The main scientific objectives of \darwin/TPF are the detection of rocky
planets similar to Earth and the spectroscopic analysis
of their atmospheres at mid-infrared wavelengths (6 - 20 $\mu$m). In addition to presenting an
advantageous star/planet contrast, this wavelength range holds
several spectral features relevant for the search of biological
activity (CO$_2$, H$_2$O, and O$_3$). The observing scenario of \darwin/TPF
consists of two phases, detection and spectral characterisation,
whose relative duration can be adjusted to optimise the scientific
return\footnote{The detection phase might not be necessary if the targets
are identified in advance by radial velocity or astrometric surveys.}. During the detection phase of the mission (nominally 2
years), \darwin/TPF will examine nearby stars for evidence of
terrestrial planets. A duration of 3 years is foreseen for the
spectroscopy phase, for a total nominal mission lifetime of 5
years. An extension to 10 years is possible and will depend on the
results obtained during the 5 first years. Such an extension could
be valuable to observe more M stars (only 10\% of the baseline
time being attributed to them), search for big planets around a
significantly larger sample of stars, and additional measurements
on the most interesting targets already studied.

The \darwin/TPF target star list has been generated from the Hipparcos
catalogue, considering several criteria: the distance ($<$ 30 pc),
the brightness ($<$ 12 V-mag), the spectral type (F, G, K, M main
sequence stars), and the multiplicity (no companions within
1$^{\prime\prime}$). The corresponding star catalogue contains 1229
single main sequence stars of which 107 are F, 235 are G, 536 are K,
and 351 are M type \citep{Kaltenegger:2008}. The survey of the \darwin/TPF
stars and the possible detection of terrestrial planets will start
a new era of comparative planetology, especially by studying the
relationship between habitability and stellar characteristics
(e.g.\ spectral type, metallicity, age), planetary system
characteristics (e.g.\ orbit), and atmospheric composition.


\section{Simulated performance} \label{sec:perfo}
\subsection{The science simulator}\label{sec:simu}

The performance predictions presented in this paper have been computed using
the \darwin{} science SIMulator developed at ESA/ESTEC (\darwinsim, \citealt{Den_Hartog:2005}). This simulator
has been subject to extensive validation the past few years and
its performance predictions were recently reconciled with a similar mission simulator
developed independently at NASA/JPL for the TPF mission \citep{Lay:2007}. The two
simulation tools have shown a very good agreement in SNR,
giving similar integration times for all the Darwin/TPF targets with a discrepancy lower than 10\%
in average \citep{Defrere:2008}. These two simulators have the same
basic purpose. For a given instrumental configuration and target catalogue,
they assess the number of terrestrial planets that can be detected in the
habitable zone of nearby main sequence stars and the number of
possible follow-up spectroscopic observations during a nominal
mission time. The duration of detection and spectroscopy phases
can be adjusted to optimise the scientific return and is nominally
set to 2 and 3 years respectively. The parameters and assumptions used are
summarized in Table~\ref{Tab:assum}.

\subsubsection{Detection phase}

The starting point of the simulations is the target star
catalogue. Given a specific interferometer architecture, the
simulator first identifies the stars which are observable from L2.
For each of these observable stars, the basic
calculation consists of an assessment of the required integration
times to achieve an user-specified SNR for broad-band detection of
a hypothetical Earth-like planet located inside the habitable zone.
The habitable zone is assumed to be located between 0.7 and 1.5\,AU
for a G2V star and is scaled with the square root of the stellar luminosity (L$^{1/2}_\star$). 
Since the location of the planet around the star is a priori
unknown, the integration time is computed from the
requirement that it should ensure the detection of a planet for at
least 90\% of the possible locations in the habitable zone.
Assuming planets uniformly distributed along habitable orbits, this requires the computation of the probability
distribution for finding a planet at a certain angular distance
from the star. For each planetary position, the maximum SNR
is computed by optimisation of the baseline length. The thermal flux of the
habitable planet is assumed to be identical to that of Earth irrespective of the
distance to the star. The exozodiacal clouds are simulated by assuming 
a nominal dust density 3 times larger than that in the solar system and
a dust sublimation temperature of 1500\,K.
Under the assumption that the exozodiacal emission
is symmetric around the target star, it will be suppressed by phase chopping,
and therefore only contributes to shot noise. The noise sources included are the shot noise contributions
from stellar leakage, local and exozodiacal clouds, and
instrumental infrared background. Instability noise
is also present and is partly mitigated by phase
chopping. A complete list can be found in Table~\ref{tab:snrs}.

After the initial integration time assessment for detection, the
targets are sorted by ascending integration time, removing from
the list the targets for which the total integration time exceeds
the total time during which they are visible from L2. Considering
a slew time for re-targeting (nominally 6 hours) and an efficiency
for the remaining observing time of 70\%, the sorted list is cut
off at the moment when the cumulative integration time exceeds the
nominal survey period. Accounting for a specific time allocation for each spectral type (10\% F, 50\% G,
30\% K and 10\% M stars), the resulting list defines the number of targets
that can be surveyed during the detection phase. The actual number of planets
found will then depend on the number of terrestrial planets present in the habitable
zone of target stars ($\eta_\oplus$).

\subsubsection{Spectroscopy phase}

The number of targets which can be characterised by spectroscopy
in a given time is computed similarly. The difference with the detection
phase is that the integration times are computed for a given position in the
habitable zone. The proper procedure would be to take into account
all possible positions for the planet in a similar way to the
detection phase but this would be far too time consuming. The
strategy is then to consider only the most likely angular separation.
Then, the total integration time is determined by the requirement
to detect the absorption lines of $O_3$, $CO_2$ and $H_2O$ to a
specified SNR. For the spectroscopy of $CO_2$ and $O_3$ (without
$H_2O$), an SNR of 5 would actually be sufficient for a secure
detection \citep{Fridlund:2005}. Considering the spectroscopy of
$H_2O$ is relatively more complex. Recent results suggest that,
using a spectral resolution greater than 20, an SNR of 10 from 7.2
to 20\,$\mu$m would be sufficient for $H_2O$, $CO_2$, and $O_3$
spectroscopy (private communication with F.~Selsis, L.~Kaltenegger
and J.~Paillet). In particular, these results suggest that the
$H_2O$ band located below 7.2\,$\mu$m, which is much more
time-consuming than the $H_2O$ band beyond 17.2 $\mu$m, could be
discarded.

Two types of spectroscopic analysis are considered: the staring
spectroscopy, where the array is kept in a position such that the
planet resides on a peak of the modulation map, and the rotating
spectroscopy, where the array keeps on rotating with respect to
the target system so that the planet moves in and out of the peaks
of the modulation map. Staring spectroscopy is more efficient in
terms of signal acquisition, but requires an accurate knowledge of
the planetary orbit. As for the detection phase, the total
spectroscopy integration time should not exceed the total time
during which the target is visible during the characterisation
phase. Accounting again for a given fraction of overhead loss, the
targets are sorted with respect to ascending integration time,
terminated where the cumulative time exceeds the length of the nominal
characterisation period. The number of planets that can be characterised is then
given with the assumption that there is one terrestrial planet in the habitable zone of
each target star ($\eta_\oplus$=1).

\begin{figure}[!t]
\centering
\includegraphics[width=9.2 cm]{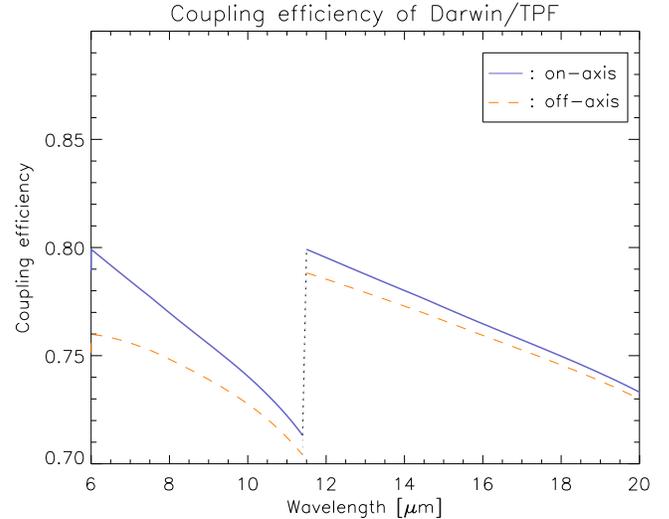}
\caption{Coupling efficiency for \darwin/TPF with
respect to the wavelength for an on-axis source and for a source with a fixed off-axis
angle, corresponding to an Earth orbit around a Sun at 10\,pc. The core radius is chosen so as to stay
single-mode on the whole wavelength.}\label{fig:coupling}
\end{figure}

\begin{figure*}[!t]
\begin{center}
\includegraphics[width=16 cm]{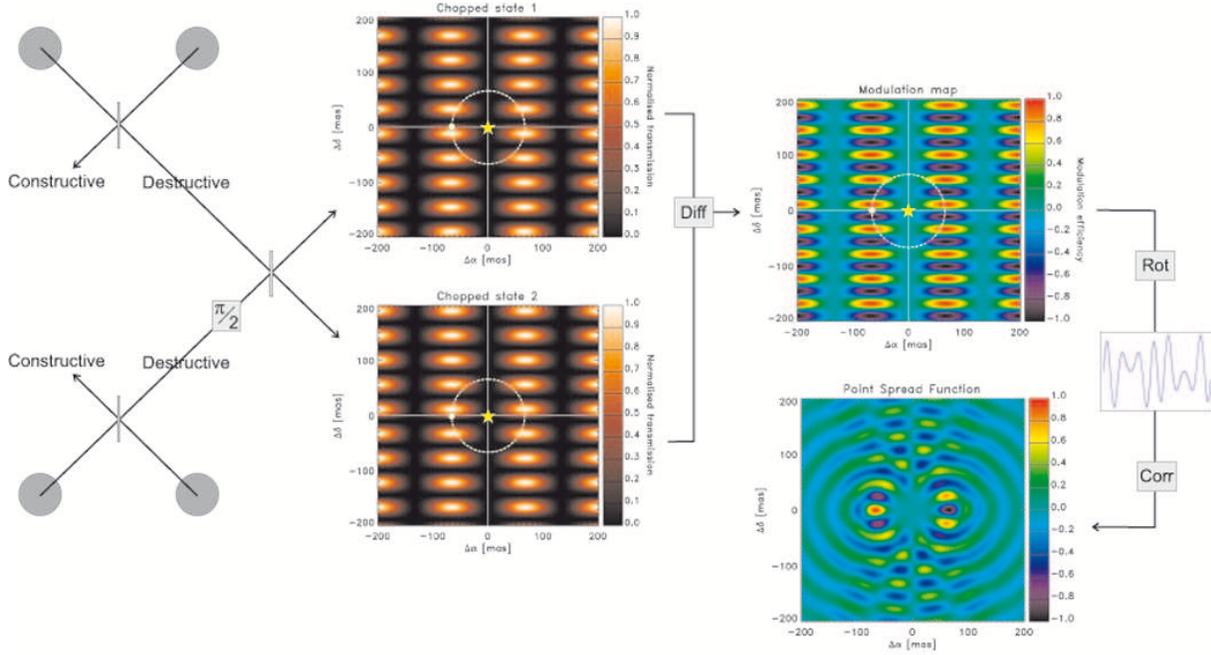}
\caption{Overview of phase chopping for the X-array configuration.
Combining the beams with different phases produces two conjugated
transmission maps (or chop states), which are used to produce
the chopped response. Array rotation then locates the planet by cross-correlation
of the modulated chopped signal with a template function.}
\label{fig:Xarray_chopping}
\end{center}
\end{figure*}

\subsection{Coupling efficiency}\label{sec:coupling}

Coupling the optical beams into optical fibers is an essential part of the
wavefront correction process, which is required for deep nulling.
The theoretical efficiency of light injection into an optical fiber
depends on several parameters: the core radius of the fiber, its numerical
aperture, the wavelength, the diameter of the telescope and its
focal length \citep{Ruilier:2001}. The method used in the simulator consists in choosing
first the core radius of the fiber so as to ensure single-mode propagation over the whole
wavelength range. The f-number of the coupling optics can then be optimised to give
the maximum coupling efficiency at a chosen wavelength and more
importantly, to provide a roughly uniformly high coupling
efficiency across the whole wavelength band. The coverage of the full
science band of \darwin/TPF with one optical fiber is generally prohibited
since the coupling efficiency drops rapidly with respect to the wavelength.
Increasing the number of fibers improves the coupling efficiency but at the expense of complexity. In particular, it has been shown
that the loss of targets for detection and spectroscopy is about 5\% between the optimised 2-band and 3-band cases \citep{Den_Hartog:2005b}.
Considering the use of two fibers (respectively on the 6.0-12 $\mu$m and 12-20 $\mu$m wavelength ranges),
Fig.~\ref{fig:coupling} shows the coupling efficiency for an on-axis source and for a source with a fixed off-axis angle, corresponding to an Earth orbit around a Sun at 10\,pc. The coupling efficiency remains above 70\% over the whole
wavelength band of each fiber.

\subsection{Modulation efficiency}

The modulation of the planetary signal during the observation is a direct consequence
of the chopping process which is mandatory to get rid of background noise sources such as exozodiacal and local zodiacal cloud emission.
For the X-array configuration, the outputs of two Bracewell interferometers are combined with
opposite phase shifts ($\pm$ $\pi/2$) to produce two transmission
maps (or ``chop states''). Differencing the two
transmission maps gives the chopped response of the interferometer, the so-called
modulation map, which contains positive and negative values by
construction (see Fig.~\ref{fig:Xarray_chopping}). Since the value of the modulation map varies across
the field-of-view, the position and flux of the planet cannot be
unambiguously inferred and an additional level of modulation is
mandatory. This is provided by the rotation of the interferometer
(typically with a period of 1 day). The planetary signal is therefore modulated
as shown on the right part of Fig.~\ref{fig:Xarray_chopping}.
In order to retrieve the planetary signal,
the most common approach is correlation mapping, a technique closely related to the Fourier
transform used for standard image synthesis \citep{Lay:2005}. The
result is a correlation map, displayed for a single point source
in the lower right part of Fig.~\ref{fig:Xarray_chopping}. This
represents the Point Spread Function (PSF) of the array. This
process, illustrated here for a single wavelength, is repeated
across the wavelength range, and the maps are co-added to obtain the net
correlation map. The broad range of wavelengths planned for
\darwin/TPF greatly extends the spatial frequency coverage of the
array, suppressing the side lobes of the PSF.

\begin{figure}[!t]
\begin{center}
\includegraphics[width=9.2 cm]{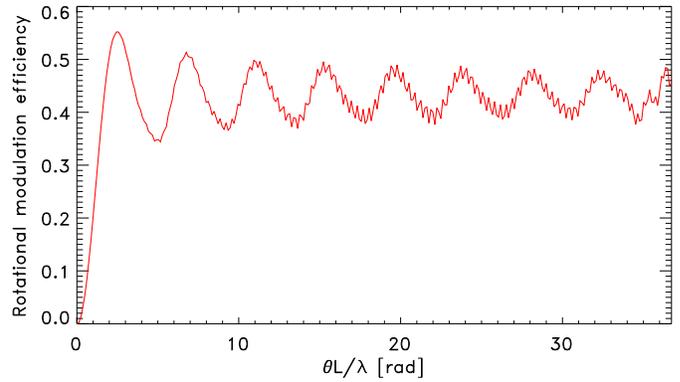}
\caption{Rotational modulation
efficiency for the Emma X-array with a 6:1 aspect ratio.} \label{fig:mod_eff}
\end{center}
\end{figure}

After chopping and rotation, the part of the incoming
signal which is actually modulated and retrievable by
synchronous demodulation is proportional to the ``rotational
modulation efficiency''. It is shown for the
X-array configuration in Fig.~\ref{fig:mod_eff}. It depends on the radial
distance from the star and reaches a peak value of 0.56 with an asymptotic value of 0.44. Since the planet
position inside the habitable zone is a priori unknown, it is
desirable that the effective modulation efficiency is as uniform
as possible across the habitable zone to avoid too many
reconfigurations of the interferometric array. Note that
the rotational modulation efficiency for several array
configurations has been investigated by \cite{Lay:2005}.

\subsection{Instability noise}\label{sec:instab}

\begin{table*}[!t]
  \centering
\caption{Detailed view of the various contributors to the noise budget, given in
photo-electrons persecond over the 6-20\,$\mu$m wavelength range for \darwin/TPF in the Emma X-array configuration.
The final SNRs are computed over a single rotation of 50000\,s and for four targets representative of the target catalogue (see main text for further information).} \label{tab:snrs}
  \begin{tabular}{l c c c c}
    \hline
    \hline
      & M0V - 05pc & K0V - 10pc & G2V - 15pc & F0V - 20pc \\
    \hline
    Optimum baseline [m]  & 25 & 22 & 22 & 11  \\
    Planetary modulation efficiency  & 0.57 & 0.58 & 0.58 & 0.58  \\
    Planetary coupling efficiency & 0.76 & 0.76 & 0.76 & 0.75  \\
    \hline
    Stellar signal [e-/s]          & 2.6 $\times$ $10^7$ & 2.1 $\times$ $10^7$ & 1.3 $\times$ $10^7$ & 2.1 $\times$ $10^7$  \\
    Planetary signal [e-/s]        & 5.52 & 1.39 & 0.62 & 0.35  \\
    Photon noise [e-/s]            & 0.459  & 0.388 & 0.358 & 0.365  \\
    - Geometric leakage [e-/s]       & 0.304 & 0.179 & 0.101 & 0.078  \\
    - Null-floor leakage [e-/s]    & 0.019 & 0.017 & 0.013 & 0.017  \\
    - 3-zodi signal [e-/s]           & 0.077 & 0.078 & 0.075 & 0.121  \\
    - Local zodiacal signal [e-/s]   & 0.260 & 0.260 & 0.260 & 0.260  \\
    - Detected thermal [e-/s]        & 0.101 & 0.101 & 0.101 & 0.101  \\
    - Detected stray light [e-/s]    & 0.166 & 0.166 & 0.166 & 0.166  \\
    Dark current [e-/s]            & 0.063 & 0.063 & 0.063 & 0.063  \\
    Detector noise [e-/s]          & 0.063 & 0.063 & 0.063 & 0.063  \\
    Instability noise [e-/s]       & 0.025 & 0.014 & 0.008  & 0.012  \\
    - First order phase term [e-/s]       & 0.021 & 0.008 & 0.002 & 0.001  \\
    - Second order phase-amplitude [e-/s] & 0.014 & 0.012 & 0.007 & 0.012  \\
    Total noise [e-/s]             & 0.460 & 0.388 & 0.358 & 0.365  \\
    \hline
    Integrated SNR (one rotation) & 16.0 & 4.4 & 2.1 & 1.1\\
    Integration time for SNR=5 [h] & 1.36 & 17.9 & 81.8 & 285\\
    \hline
  \end{tabular}
\end{table*}

Instability noise is defined as the component of the demodulated
nulled signal that arises from phase, amplitude and polarisation errors \citep{Lay:2004}.
The power spectra of these instrumental effects mix with each other so that perturbations at all frequencies,
including DC, have an effect. Spacecraft vibrations, fringe tracking offset, control noise, longitudinal chromatic
dispersion, and birefringence are at the origin of the phase errors whereas tip/tilt, defocus, beam shear, and differential transmission
produce amplitude errors. These phase and amplitude errors induce
a time-dependent asymmetry between the two chop states so that the modulation map does not remain centered on the
nominal position of the line of sight (i.e.\ the position of the
star). Hence a fraction of the starlight survives the modulation
process and mixes with the planet photons. Although a simple binary
phase chop removes a number of these systematic errors, it has no effect on the dominant
amplitude-phase cross terms and on the co-phasing errors. There is no phase chopping
scheme that can remove the systematic errors without also removing the planetary signal.

Three independent studies \citep{Lay:2004,Darcio:2005,Chazelas:2006} have reviewed the
instrumental requirements on the \darwin/TPF mission that reduce the instrumental stellar leakage 
to a sufficiently low level for Earth-like planet detection. Assuming the presence of
1/f-type noise, these studies showed that the requirements on
amplitude and phase control are not driven by the null-floor
leakage, but by instability noise. Considering a Dual-Chopped Bracewell (DCB, \citealt{Lay:2004})
with 4-m aperture telescopes operating at 10\,$\mu$m, the
different analyses show that a null depth of $\sim$ 10$^{-5}$ is
generally sufficient to control the level of shot noise from the
stellar leakage, but that a null depth of $\sim$ 10$^{-6}$ is
required to prevent instability noise from becoming the dominant
source of noise\footnote{The current state-of-the-art for broadband nulling experiments is a 10$^{-5}$ null which
has been recently demonstrated at 10\,$\mu$m ($\Delta\lambda$/lambda=34\%) with the adaptive
nuller \citep{Peters:2009}.}. In particular, a 10$^{-6}$ null requires rms path control to
within about 1.5\,nm, and rms amplitude control of about 0.1\%.

In order to relax these very stringent requirements, several
techniques have been investigated \citep{Lay:2005,Lane:2006,Gabor:2008}.
Discussion of these mitigation techniques is beyond the scope of this paper, where we assume that instability noise is
sufficiently low to ensure the H$_2$O spectroscopy at 7-$\mu$m of an Earth-like planet
orbiting around a Sun located at 15\,pc. Applying the analytical method of \citep{Lay:2004} to the Emma X-array with the parameters
listed in Table~\ref{Tab:darwin_spec}, we derive the constraints on the instrument stability such
that instability noise is dominated by shot noise by a factor 5 at 7 $\mu$m over one rotation
period ($t_{\rm rot}$) of 50000 seconds (with a spectral resolution of 20). Considering 1/f-type PSDs defined on
the [1/$t_{\rm rot}$,10$^4$]\,Hz range, this corresponds to residuals rms OPD and amplitude errors of about 1.5\,nm
and 0.05\% respectively. These values will be used thorough this study (see appendix~\ref{app:instab} for further details).
Although our computation has been done at 7\,$\mu$m where instability noise is much higher than at 10\,$\mu$m, these constraints
are not far from the values derived by \cite{Lay:2004} for two reasons. First, the telescopes considered here are smaller so that
shot noise is relatively more dominant than in the previous analyses (shot noise is proportional to the square root of the stellar flux
while the planetary signal and instability noise are directly proportional to the stellar flux). Secondly, the interferometer configuration
is stretched by comparison with the DCB so that the planetary signal is modulated at higher frequencies, where
the instability noise is lower assuming 1/f-type PSDs.


\subsection{Signal-to-noise analysis}\label{sec:snr}

In this section, we present the different sources of noise
simulated by \darwinsim{} and the level at which they contribute to
the final SNR for four targets representative of the \darwin/TPF catalogue: an M0V star located at 5\,pc,
a K0V star at 10\,pc, a G2V star at 15\,pc and an F0V star at 20\,pc. The noise budget of each source
is shown in Table~\ref{tab:snrs} for a single rotation of 50000\,s and for the optimum baseline length (computed
by minimizing the integration time). The different contributors are described hereafter.

\begin{itemize}
\item The stellar signal represents the total number of photo-electrons that are generated by stellar photons
detected in both constructive and destructive outputs.
\item The planetary signal is the demodulated amount of photo-electrons
that come from an Earth-like planet located at 1\,AU from the star.
\item Shot noise is due to the statistical arrival process of the photons from all sources. It comes from the contributions from stellar leakage, the exozodiacal dust, the local zodiacal cloud, the thermal emission from the telescopes, and the stray light.
\item Geometric stellar leakage accounts for the imperfect rejection of the stellar photons due to the
finite size of the star and the non-null response of the interferometer for small off-axis angles.
\item Null-floor leakage accounts for the stellar photons that leak through the output of the interferometer due to the influence of instrumental
imperfections such as co-phasing errors, wavefront errors or mismatches in the intensities of the beams.
\item The 3-zodi signal is the shot noise contribution from the circumstellar disc, assumed to
be face-on and to follow the same model as in the solar system \citep{Kelsall:1998}, except for a global density factor of 3.
\item Local zodiacal signal is the shot noise contribution from the solar zodiacal cloud, taking into account the spacecraft location at L2 and the pointing direction.
\item Thermal background accounts for the emission of the telescopes.
\item Stray light is made of the photons originating from outside the interferometer and which do not follow the nominal route to the detector. It includes scattered light from the target star, thermal photons from the instrument and any solar photon that are scattered into the instrument. We assume a nominal value of 10 photons per second and per spectral channel.
\item Dark current is the constant response produced by the detector when it is not actively being exposed to light. We consider a nominal value of of 4 electrons rms per read and per spectral channel.
\item Detector noise is computed assuming a read-out noise of 4 electrons rms and a typical read-out frequency of 1\,Hz.
\item Instability noise has been discussed in section~\ref{sec:instab}. It is computed for rms OPD and amplitude errors of 1.5\,nm and 0.05\% respectively (and defined on 1/f-type power spectra).
\end{itemize}

\begin{figure}[!t]
\begin{center}
\includegraphics[width=9.2 cm]{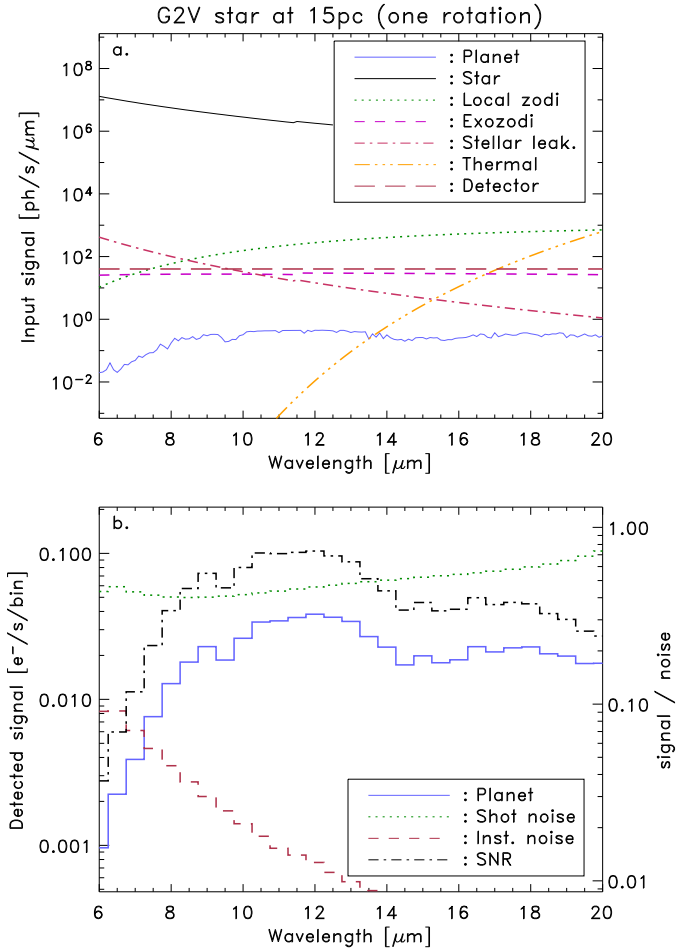}
\caption{Input (a) and detected (b) signals for an Earth-like planet
orbiting at 1\,AU around a G2V star located at 15\,pc. The demodulated signals
are computed over a single rotation of 50000\,s.} \label{fig:noise_wav}
\end{center}
\end{figure}

As expected, shot noise is the dominant contributor for all targets (we have derived the constraints on the instrument stability so that instability noise is not dominant in section \ref{sec:instab}). The emission of the local zodiacal cloud is generally the dominant source of shot noise but geometric stellar leakage and the exozodiacal dust also produce a significant part of it. Instability noise is dominated by second order phase-amplitude cross terms as pointed out by \cite{Lay:2004} and contributes weakly to the integrated SNR. The dependence on the wavelength is shown in Fig.~\ref{fig:noise_wav} for the G2V star located at 15\,pc. The upper figure represents the different input signals, showing that stellar leakage dominates at short wavelengths while the local zodiacal cloud emission is dominant at long wavelengths. The lower figure represents shot noise, instability noise and the detected signal from the planet. The SNR is also represented and measured on the right-hand side vertical axis. It is maximum around 10\,$\mu$m, where the planetary signal peaks, and decreases rapidly toward short wavelengths, suggesting that the spectroscopy at these wavelengths would be the most time-consuming. Integrating the signal and noise sources from 6 to 20\,$\mu$m gives an SNR of about 2 (see table~\ref{tab:snrs}) so that about 6 rotations of the interferometric array would be necessary to achieve the detection of the planet with an SNR of 5. For spectroscopy, the integration time has to be much longer since the individual SNR in each spectral channel is significantly lower. For instance, about 150 rotations of the interferometric array would be necessary to achieve an SNR of 10 at 10\,$\mu$m.

\begin{table}[!t]
\begin{center}
\caption{Expected performance in terms of number of stars surveyed
and planets characterised during the nominal 5-year mission for
various telescope diameters and planet radii. All stars are assumed to host a planet in the
habitable zone and to be surrounded by an exozodiacal cloud 3
times denser that in the solar system.}\label{tab:darwin_results}
\begin{tabular}{l c c c | c c}
\hline \hline
  Telescope diameter & 1-m & 2-m & 4-m & 2-m & 2-m \\
  Planet radius [$R_\oplus$] & 1 & 1 & 1 & 1.5 & 2 \\
  \hline
  \textbf{Detection}  &  &  &  &  &  \\
  Surveyed (5 years) & 89 & 303 & 813 & 590 & 921 \\
  Surveyed (2 years) & 58 & 189 & 497 & 370 & 564 \\
  \hspace{0.3 cm} \# F stars & 3 & 10 & 35 & 27 & 46 \\
  \hspace{0.3 cm} \# G stars & 11 & 43 & 136 & 96 & 164 \\
  \hspace{0.3 cm} \# K stars & 14 & 61 & 183 & 130 & 206 \\
  \hspace{0.3 cm} \# M stars & 30 & 75 & 143 & 117 & 148 \\
  \hline
  \textbf{Spectroscopy}  &  &  &  &  &  \\
  Staring ($CO_2$,$O_3$) & 20 & 64 & 199 & 132 & 234 \\
  Rotating ($CO_2$,$O_3$) & 20 & 43 & 127 & 89 & 159 \\
  \hspace{0.3 cm} \# F stars & 0 & 2 & 5 & 4 & 8 \\
  \hspace{0.3 cm} \# G stars & 2 & 7 & 25 & 17 & 36 \\
  \hspace{0.3 cm} \# K stars & 2 & 10 & 40 & 24 & 46\\
  \hspace{0.3 cm} \# M stars & 16 & 24 & 67 & 44 & 69 \\
  Staring ($H_2O$) & 15 & 32 & 101 & 71 & 121 \\
  Rotating ($H_2O$) & 11 & 21 & 60 & 48 & 83 \\
  \hspace{0.3 cm} \# F stars & 0 & 1 & 2 & 2 & 4 \\
  \hspace{0.3 cm} \# G stars & 0 & 4 & 11 & 8 & 16 \\
  \hspace{0.3 cm} \# K stars & 2 & 5 & 18 & 11 & 22 \\
  \hspace{0.3 cm} \# M stars & 9 & 11 & 39 & 27 & 41 \\
  \hline
\end{tabular}
\end{center}
\end{table}

\subsection{Expected performance}

Considering the assumptions given in Table~\ref{Tab:assum}, the
simulated performance of \darwin/TPF is shown in
Table~\ref{tab:darwin_results} for various aperture sizes and planet radii.
Considering Earth-radius planets within the habitable zone, about 200 stars, well spread
among the four selected spectral types, can be surveyed during the
nominal 2-year detection phase. This number reaches about 500 with 4-m
aperture telescopes. \darwin/TPF will thus provide statistically
meaningful results on nearby planetary systems.
As already indicated by Table\,\ref{tab:snrs}, nearby K and M dwarfs
are the best-suited targets in terms of Earth-like planet detection
capabilities.

For the spectroscopy phase, a required SNR of 5 has been assumed
for the detection of $CO_2$ and $O_3$, as discussed in section
\ref{sec:simu}. For the full characterisation (i.e.\
searching for the presence of $H_2O$, $CO_2$, and $O_3$), the
required SNR has been fixed to 10 on the 7.2-20-$\mu$m wavelength
range. With these assumptions, $CO_2$ and $O_3$ could be searched
for about 40 planets (resp.~60) with rotational spectroscopy
(resp.~staring spectroscopy) while $H_2O$ could potentially be
detected on 20 (resp.~30) planets during the 3-year
characterisation phase. These values would be roughly halved for 1-m
aperture telescopes. Although staring spectroscopy presents (as expected)
better results, rotational spectroscopy is more secure since it
does not rely on an accurate localisation of the planet. It is
also interesting to note that in the case of planets with radii
1.5 time as large as that of Earth, the number of
planets for which $H_2O$ spectroscopy could be performed is
doubled.

\begin{figure}[!t]
\begin{center}
\includegraphics[width=8.5 cm]{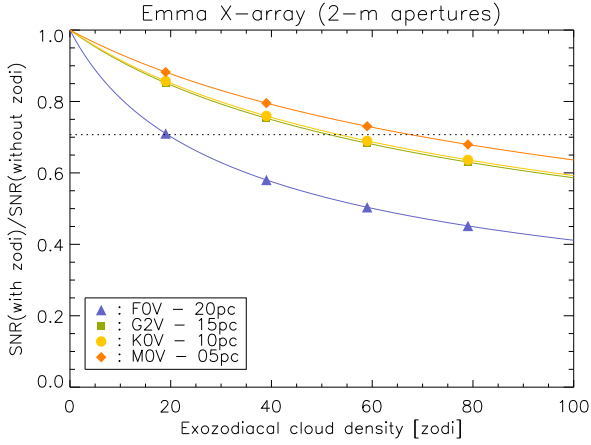}
\caption{Impact of the exozodiacal dust density on the SNR for different target stars.
The exozodiacal disc is assumed to follow the Kelsall model \citep{Kelsall:1998} and
to be seen in face-on orientation. The horizontal dotted line corresponds to an increase of integration time
by a factor 2 with respect to the 0-zodi case.} \label{fig:SNR_vs_zodi}
\end{center}
\end{figure}

\section{Impact of the exozodiacal cloud density}\label{sec:disc_dens}

The amount of exozodiacal dust in the habitable zone
of nearby main sequence stars is one of the main design drivers
for the \darwin/TPF mission. Depending on their morphology and
brightness, exozodiacal dust clouds can seriously hamper the
capability of a nulling interferometer to detect and characterise
habitable terrestrial planets. Under the assumption that it is centrally
symmetric around the target star, the exozodiacal cloud is suppressed by the
chopping process, and therefore only contributes to shot noise.
An exozodiacal cloud similar to the local zodiacal disc
emits 350 times as much flux at 10\,$\mu$m than an Earth-like planet, so that
it generally drives the integration time as the disc becomes a few times denser than the
local zodiacal cloud. A previous study performed for the DCB with 3-m aperture telescopes
observing a G2V star located at 10\,pc has led to the conclusion that
detecting Earth-like planets around a star for which the exozodiacal
cloud density is larger than 20 zodis would be difficult \citep{Beichman:2006b}.
Nevertheless, the tolerable amount of dust around a nearby main sequence star highly depends on several parameters
such as the telescope size, the target distance and spectral type.
Another parameter which can affect the performance of the interferometer is the presence
of asymmetric structures in the exozodiacal disc such as clumps or offset due
to the presence of planets. These asymmetries have a different impact on the mission performance
because they introduce a signal which is not perfectly suppressed by phase chopping and can mimic
the planetary signal. This section is focused on the impact of the exozodiacal dust density on the
integration time and the consequence on the number of targets that can be surveyed during the mission lifetime.
The impact of asymmetric structures is discussed in section~\ref{sec:disc_morph}. 

\subsection{Analysis per individual target}

\begin{figure}[!t]
\begin{center}
\includegraphics[width=8.5 cm]{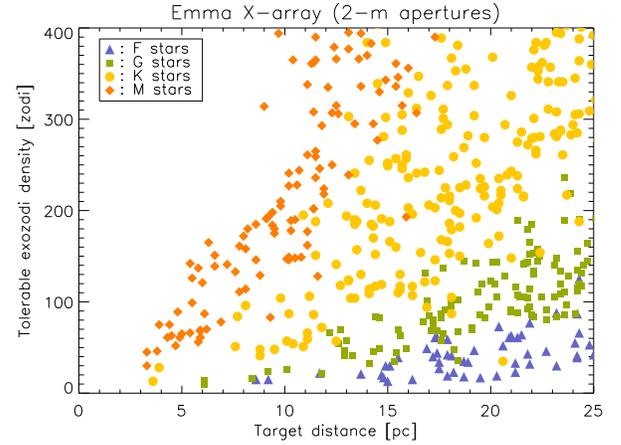}
\caption{Maximum number of zodis with respect to the target distance for the \darwin/TPF target stars. The maximum number of zodis corresponds to an increase of integration time by a factor two with respect to the 0-zodi case.}\label{Fig:zodi_vs_dist}
\end{center}
\end{figure}

Currently, very little is known about the amount of exozodiacal dust in the habitable zone of nearby main sequence stars. First results have been obtained only very recently using classical infrared interferometry at the CHARA array (Mount Wilson, USA) and at the VLTI (Cerro Paranal, Chile). These instruments have revealed the presence of hot dust in the inner part of planetary systems around a few nearby main sequence stars with a sensitivity of approximately one thousand zodis \citep{Absil:2006b,Difolco:2007,Absil:2007b,Absil:2009,Akeson:2009}. Recent observations using ground-based nulling interferometry at the Keck observatory (Mauna Kea, USA) have shown improved sensitivity to exozodiacal dust clouds of a few hundred zodis \citep{Stark:2009b}. Given the lack of information on exozodiacal clouds with densities of a few zodis, we investigate in this section the impact of exozodiacal dust density on the performance of \darwin/TPF.

Considering centrally symmetric face-on exozodiacal discs, Fig.~\ref{fig:SNR_vs_zodi} shows the SNR (normalised to the SNR for no exozodiacal cloud) integrated over the 6-20\,$\mu$m wavelength range as a function of the exozodiacal dust density for the 4 typical target stars used in section~\ref{sec:snr}. We consider the normalised SNR because it does not depend on the integration time (the planetary signal is removed from the equation) which has the advantage to provide a common basis to compare the different target stars. Looking at Fig.~\ref{fig:SNR_vs_zodi}, the impact of the exozodiacal dust density on the normalised SNR is particularly harmful for the hottest target stars which present the brightest exozodiacal discs. For the F0V star located at 20\,pc, the normalised SNR is reduced by a factor of about 2.5 between the 0 and 100-zodi cases while for the M0V star located at 5\,pc it is only reduced by a factor of about 1.5. Assuming that the integration time should not be twice longer than in the 0-zodi case (see the horizontal dotted curve), the maximum number of zodis are about 70, 50, 50 and 20 respectively for the M0V star located at 5\,pc, the K0V star at 10\,pc, the G2V star at 15\,pc and the F0V star at 20\,pc.

The distance of the star also plays an important role. As the distance to the target system increases, the flux collected from the exozodiacal cloud decreases while the flux collected from the local zodiacal cloud remains the same for all targets. The contribution of the exozodiacal dust cloud to the noise level becomes therefore relatively less important so that a higher dust density can be tolerated around the target. This explains why the curves of the G2V star and the K0V star almost coincide despite the fact that the G2V star is hotter. This behavior is illustrated in Fig.~\ref{Fig:zodi_vs_dist}, showing the maximum number of zodis with respect to the target distance for the whole \darwin/TPF catalogue. This maximum number of zodis corresponds to an increase of integration time by a factor two with respect to the 0-zodi case. It depends on the target distance and spectral type, and can take a value from few zodis up to several hundred zodis for the most distant stars. For a given distance to the target system, the maximum number of zodis increases with the stellar temperature, the zodi constraint being more severe on F stars than on M stars, while for a given spectral type, the zodi tolerance increases with the target distance.


\subsection{Tolerable dust density}

\begin{figure}[!t]
\begin{center}
\includegraphics[width=8.5 cm]{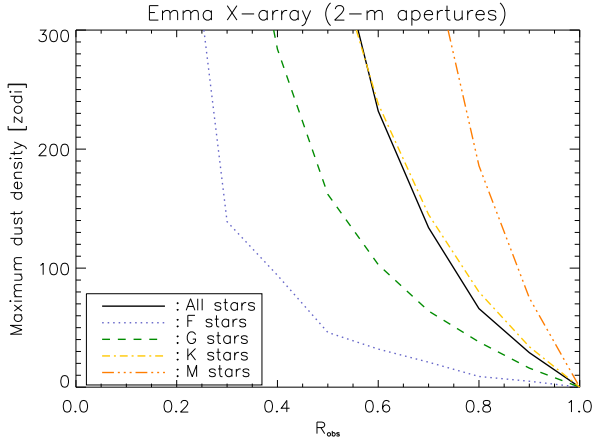}
\caption{Tolerable exozodiacal dust density as a function of the number of targets that can be observed during the mission lifetime normalised to the zodi-free case (R$_{\rm obs}$), accounting for overheads and a specific time allocation for each spectral type (10\% F, 50\% G, 30\% K and 10\% M).}\label{Fig:zodi_vs_ratio}
\end{center}
\end{figure}

So far we have examined the maximum exozodiacal dust density for each target which corresponds to an integration time per target equal to twice the zero-zodi integration time.  However, this does not tell us anything about the total number of stars that could be observed over the survey time of the mission. Here we derive the exozodiacal dust density that can be tolerated around nearby main sequence stars so that a given number of stars can be observed during the nominal mission lifetime.

To calculate the total number of targets that can be surveyed during the mission lifetime, we compute the integration time for each observable target as a function of the exozodiacal dust density and add them in ascending order as described in section~\ref{sec:simu}. Considering a slew time of 6 hours and an efficiency for the remaining observing time of 70\%, the list is cut off when the cumulative integration time exceeds the nominal survey period. Applying this procedure to each spectral type (with the corresponding time allocation, see Table~\ref{Tab:assum}) and to the whole target list, Fig.~\ref{Fig:zodi_vs_ratio} shows the tolerable dust density as a function of R$_{\rm obs}$, the ratio of the number of targets that can be observed during the mission lifetime in the presence of exozodiacal clouds of the given density  to the number of targets that can be observed during the mission lifetime in the absence of exozodiacal clouds. The corresponding number of target stars that can be observed can easily be computed using Table~\ref{tab:darwin_results}.

As exozodiacal discs become denser, R$_{\rm obs}$ decreases so that the tolerable dust density depends on the goal of the mission in terms of the number of stars that have to be observed. In order to survey approximately 50\% of the stars that could be observed in the absence of exozodiacal discs, a dust density as high as 400\,zodis can be tolerated, while only 30\,zodis are tolerable to observe 90\% of the stars. Considering that at least 150 targets\footnote{Detecting 150 targets has been defined by both ESA and NASA as the minimum mission requirement for \darwin/TPF-I.} have to be observed during the mission lifetime (about 75\% of the stars), exozodiacal discs with a density of about 100\,zodis can be tolerated around the targets stars. This tolerable dust density is computed for the whole catalogue assuming that the time allocation on each spectral type is maintained. In practice, the effect of exozodiacal dust is more pronounced for early type stars. This behavior is illustrated in Fig.~\ref{Fig:zodi_vs_ratio} for each spectral type. For a dust density of 100\,zodis, about 40\%, 60\%, 80\% and 90\% of the F, G, K and M stars respectively can still be surveyed. Conversely, in order to survey at least 75\% of the stars, the tolerable dust densities are about 10, 50, 100 and 300 \,zodis for F, G, K and M stars respectively.

These results show how important it is to observe in advance the \darwin/TPF targets in order to maximize the number of stars that can be surveyed during the mission lifetime. Ground-based nulling instruments like LBTI (``Large Binocular Telescope Interferometer", \citealt{Hinz:2008b}) and ALADDIN (``Antarctic L-band Astrophysics Discovery Demonstrator for Interferometric Nulling", \citealt{Absil:2008}) would be ideal to reach the detection of 50-zodi exozodiacal discs with a sky coverage sufficient to observe almost the entire \darwin/TPF catalogue.

\subsection{Influence of the telescope size}

\begin{figure}[!t]
\begin{center}
\includegraphics[width=8.5 cm]{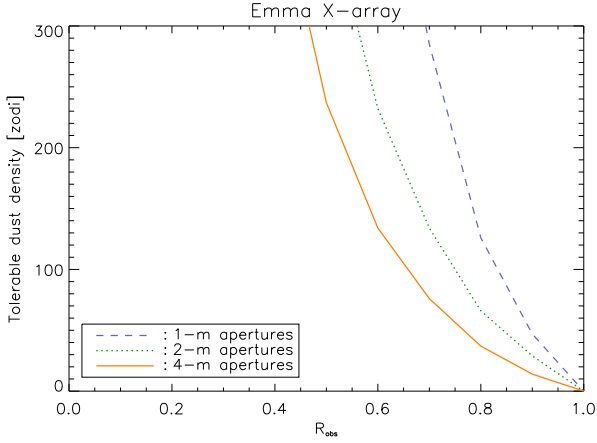}
\caption{Tolerable exozodiacal dust density for different aperture sizes as a function of the number of targets that can be observed during the mission lifetime normalised to the zodi-free case (R$_{\rm obs}$). The results are computed for the nominal mission lifetime accounting for overheads and a specific time allocation for each spectral type (10\% F, 50\% G, 30\% K and 10\% M). }\label{Fig:zodi_vs_diam}
\end{center}
\end{figure}

Increasing the telescope diameter has different influences on the individual signal and noise sources. The planetary signal increases as D$^2$ while the shot noise contributions from geometric leakage and exozodiacal cloud increase as D. The relative contribution from the local zodiacal cloud to shot noise is reduced for larger aperture telescopes, due to the smaller field-of-view. Since the local zodiacal cloud emission is generally one of the dominant noise sources, the relative impact of the exozodiacal cloud density on the SNR becomes therefore more significant for larger telescopes.
Using the assumptions of Table~\ref{Tab:assum}, this behavior is illustrated in Fig.~\ref{Fig:zodi_vs_diam}, which shows the tolerable
exozodiacal dust density with respect to R$_{\rm obs}$ for different aperture sizes.

For a given dust density, the loss of observable targets during the mission lifetime with respect to the case without exozodiacal disc is more important for larger aperture telescopes. For instance, the tolerable densities are 50, 30 and 15\,zodis respectively for 1-m, 2-m and 4-m apertures telescopes in order to survey at least 90\% of the nominal targets. These values become 200, 100 and 60\,zodis if 75\% of the targets have to be surveyed. Considering again that 150 targets have to be observed during the mission lifetime, dust densities as high as 100 and 600\,zodis can be tolerated around the target stars for 2-m and 4-m apertures telescopes respectively (150 targets being not detectable within the survey time with 1-m aperture telescopes). Due to the better nominal performance achieved with 4-m aperture telescopes (about 500 targets surveyed during the survey time, see Table~\ref{tab:darwin_results}), the maximum dust density to survey 150 targets (600 zodis) is higher than for 2-m aperture telescopes (100 zodis) but the corresponding loss of surveyed targets with respect to the nominal case is much more important (R$_{\rm obs}$ of 30\% vs 75\%). In practice, larger aperture telescopes would obviously be better to maximize the scientific performance but the final choice will also result from a trade-off with cost and feasibility.

%
%

\section{Impact of the exozodiacal cloud morphology}\label{sec:disc_morph}

\begin{figure*}[!t]
\begin{center}
\hspace{-6mm}
\includegraphics[width=4.2 cm]{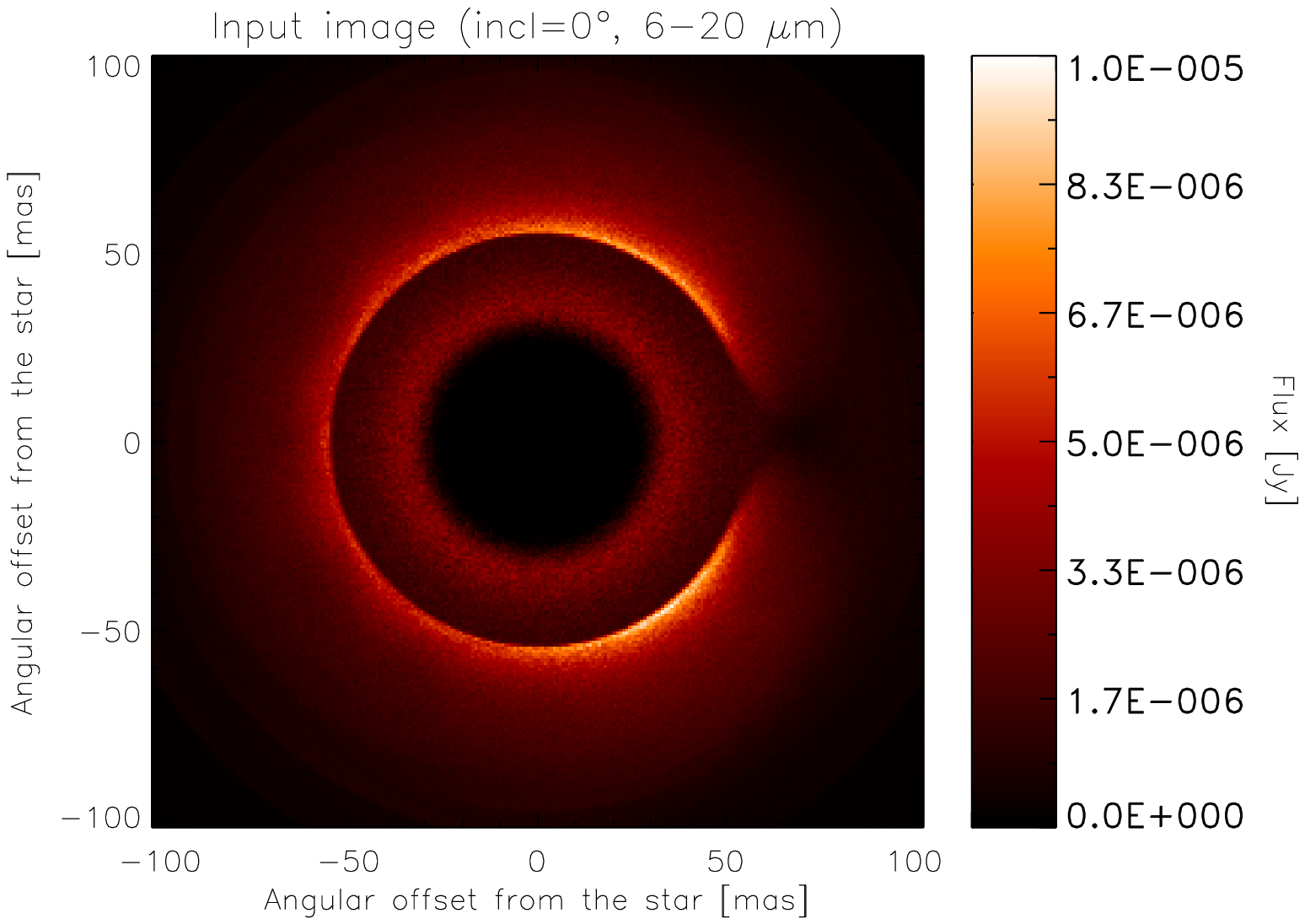}\hspace{4mm}
\includegraphics[width=4.2 cm]{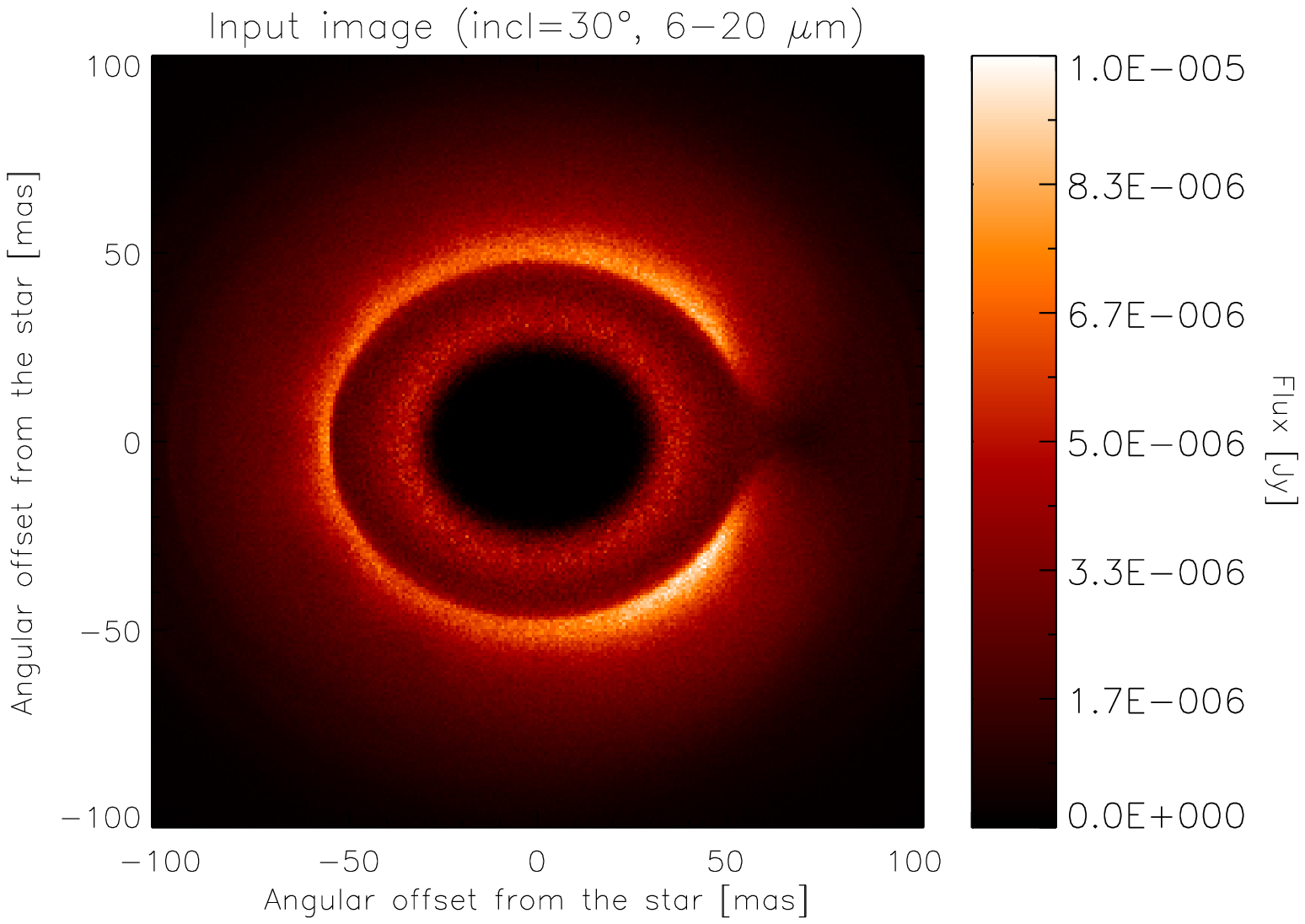}\hspace{4mm}
\includegraphics[width=4.2 cm]{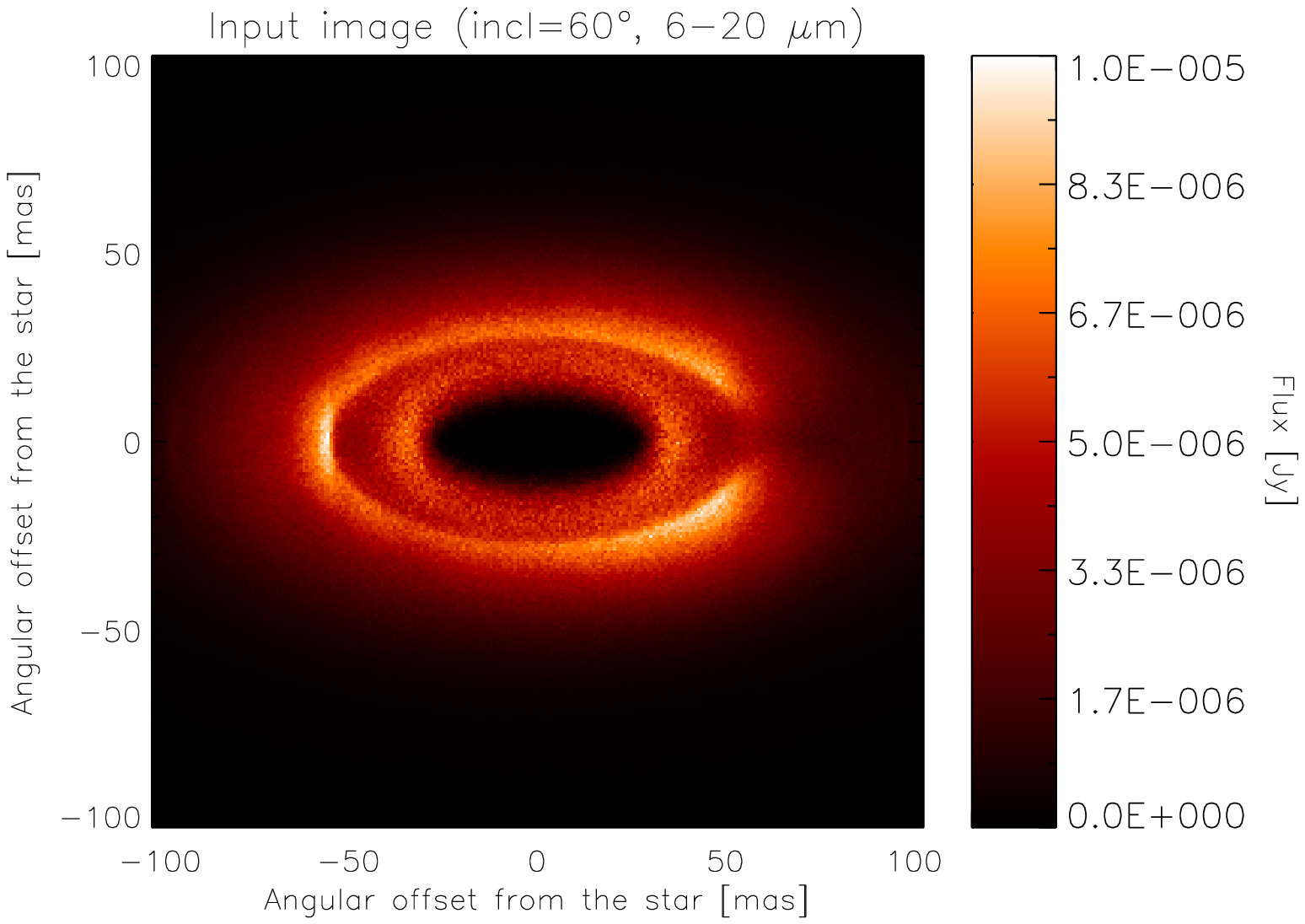}\hspace{4mm}
\includegraphics[width=4.2 cm]{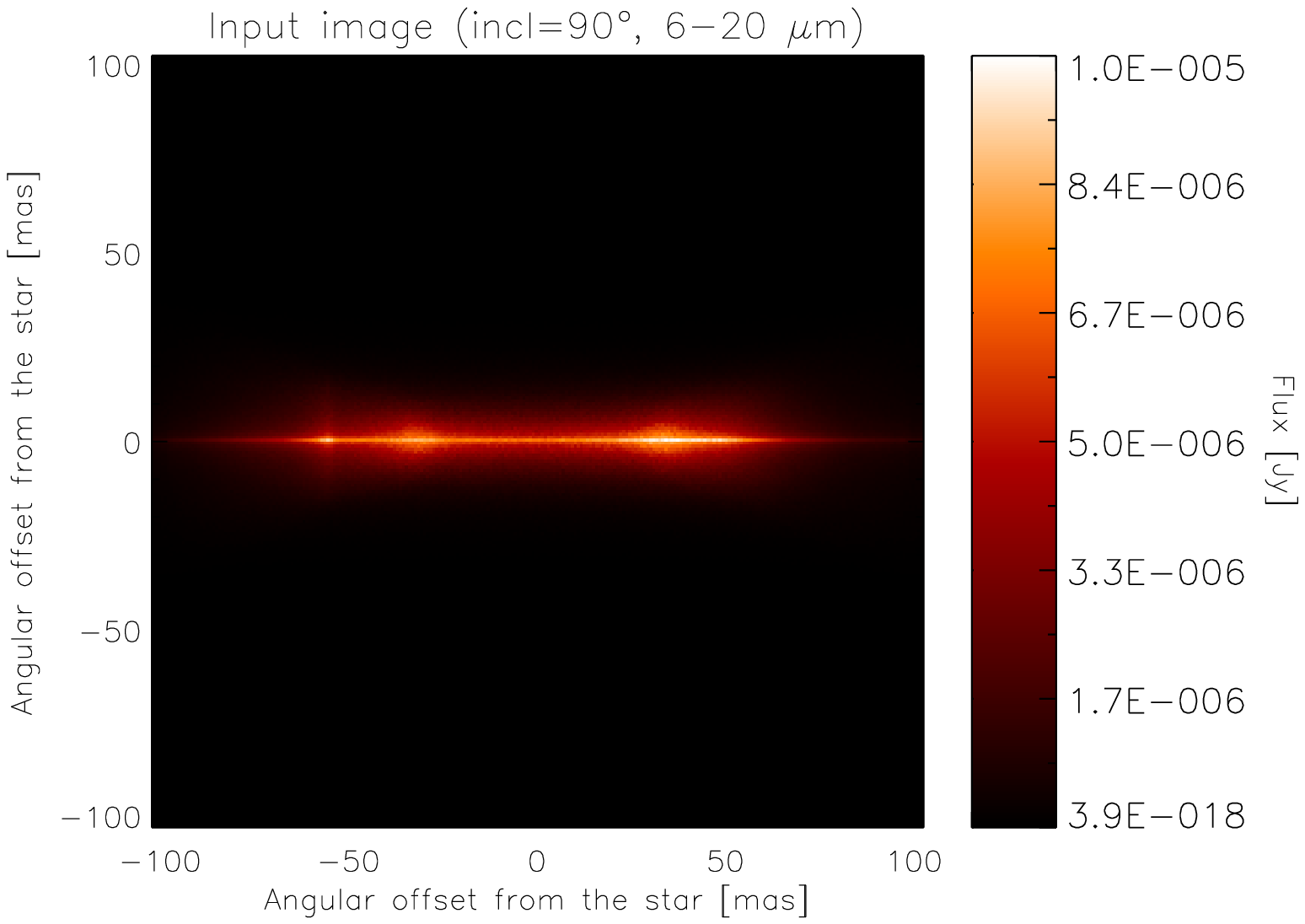}\hspace{8mm}\\
\hspace{-6mm}
\includegraphics[width=4.2 cm]{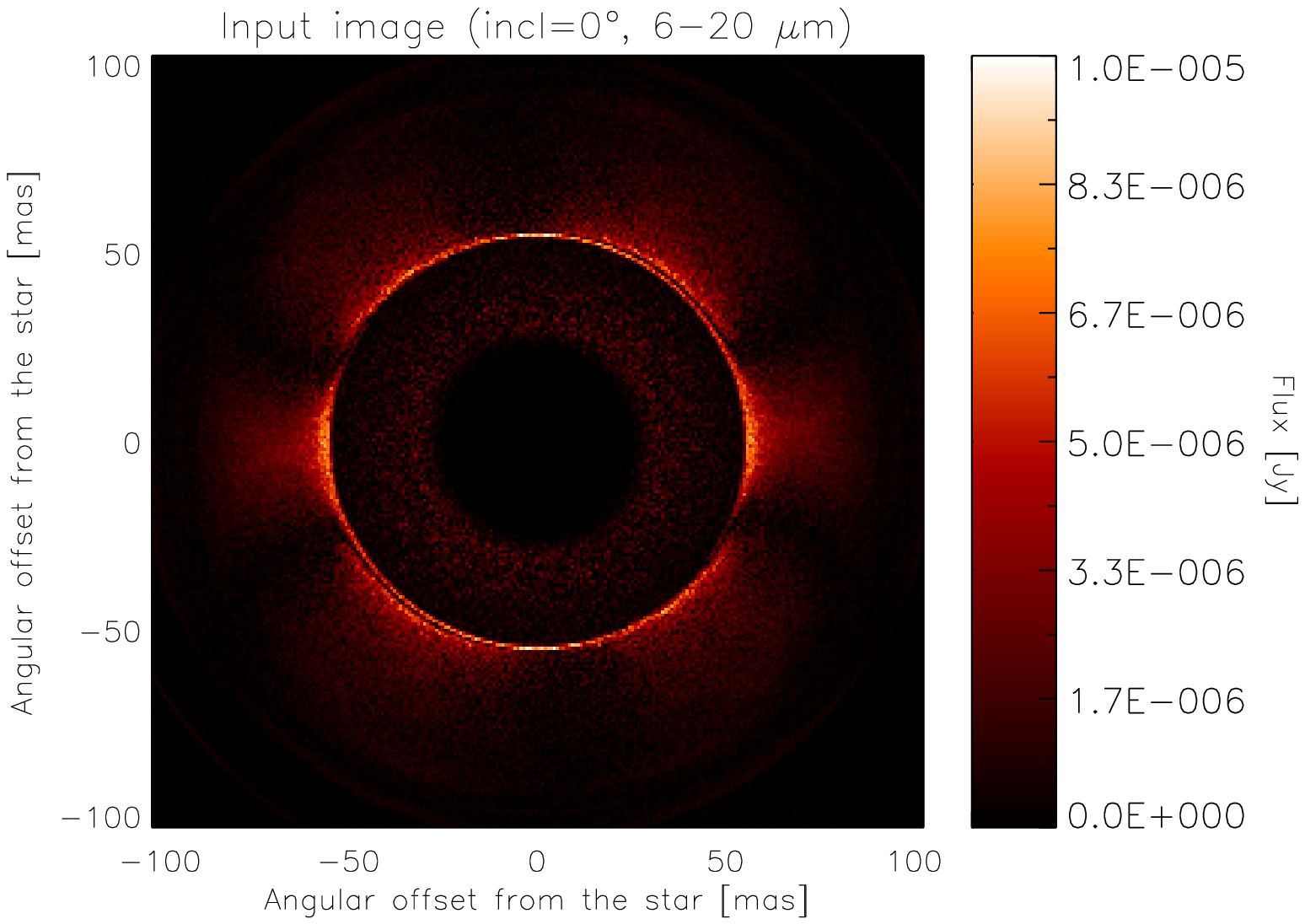}\hspace{4mm}
\includegraphics[width=4.2 cm]{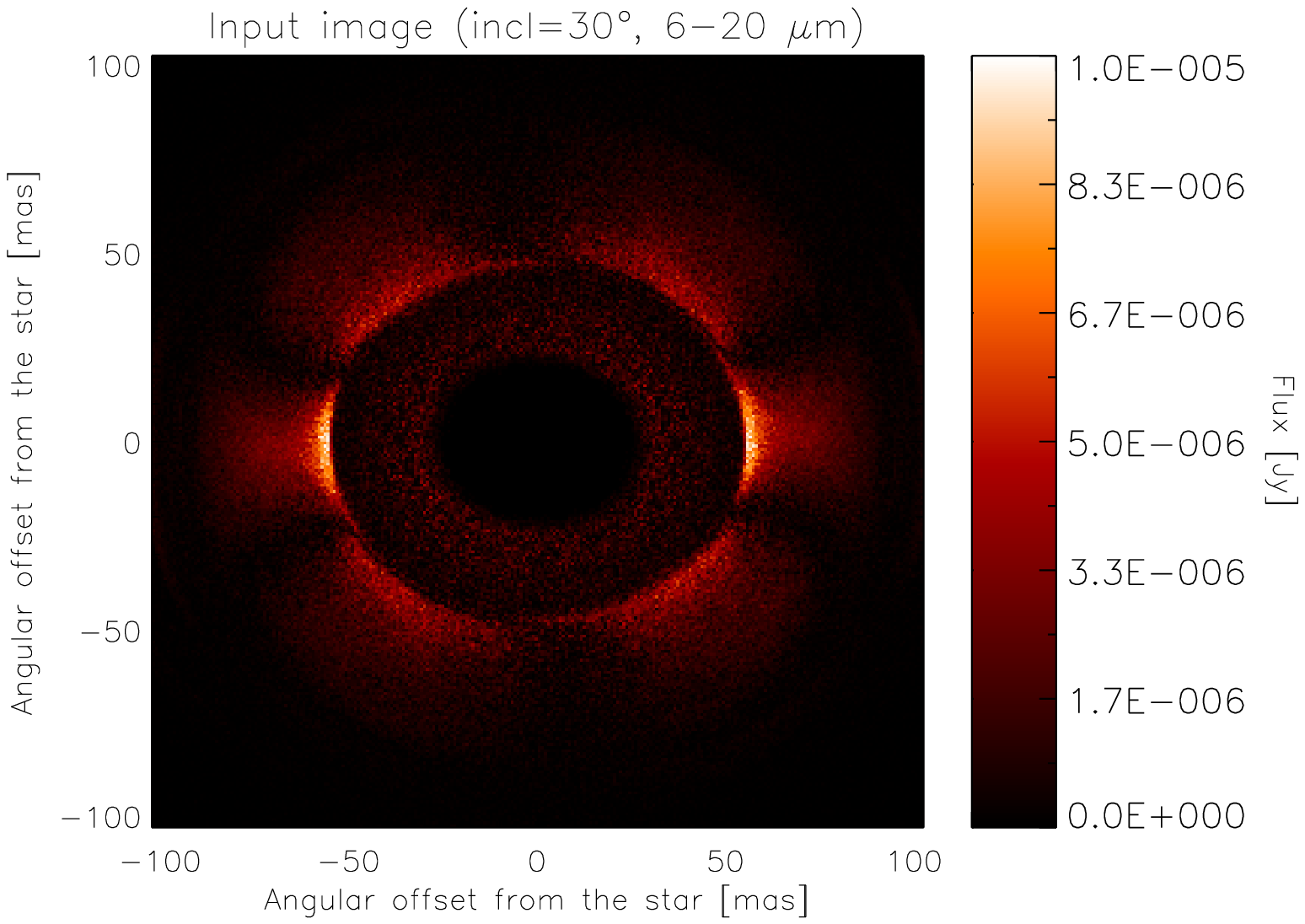}\hspace{4mm}
\includegraphics[width=4.2 cm]{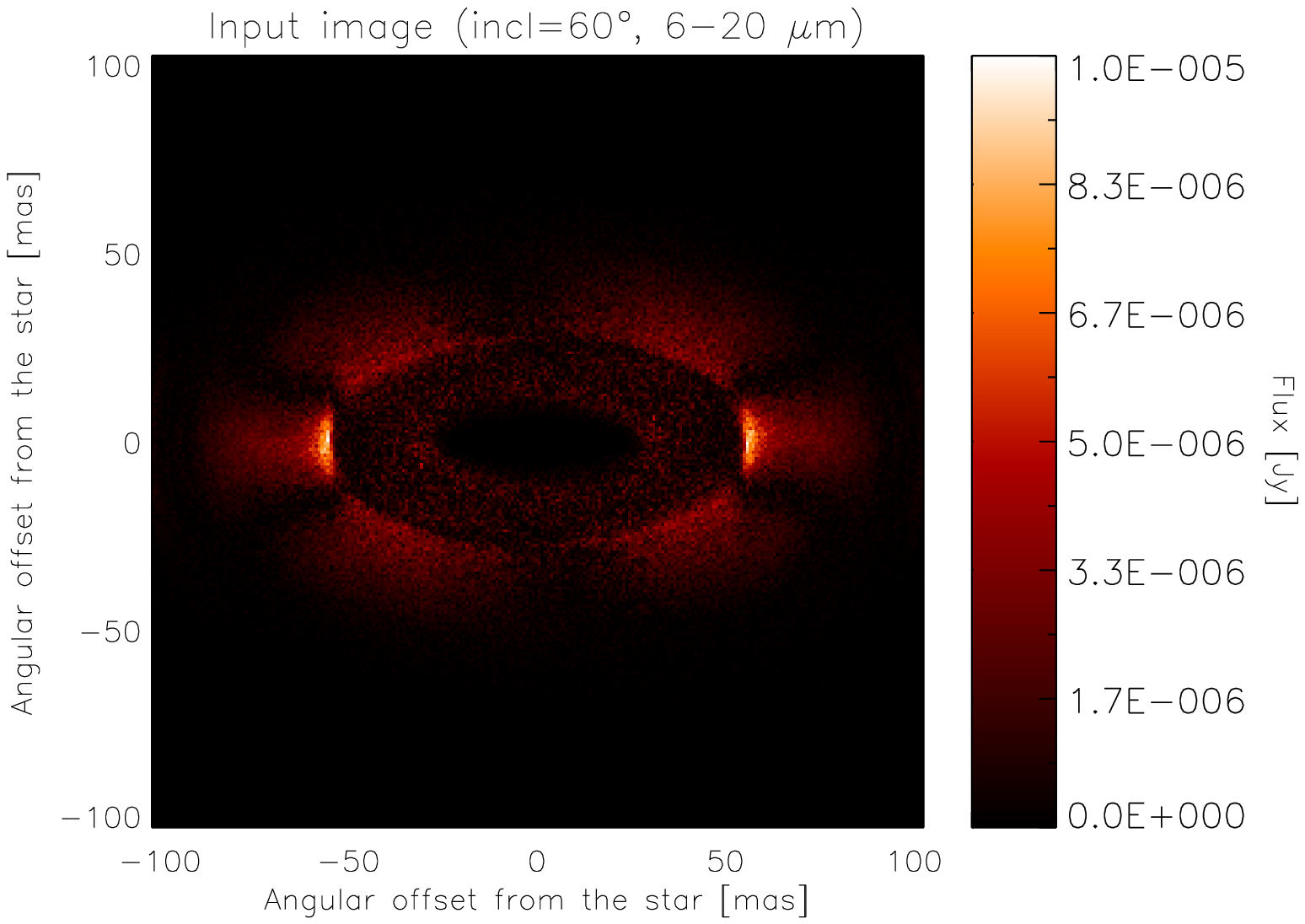}\hspace{4mm}
\includegraphics[width=4.2 cm]{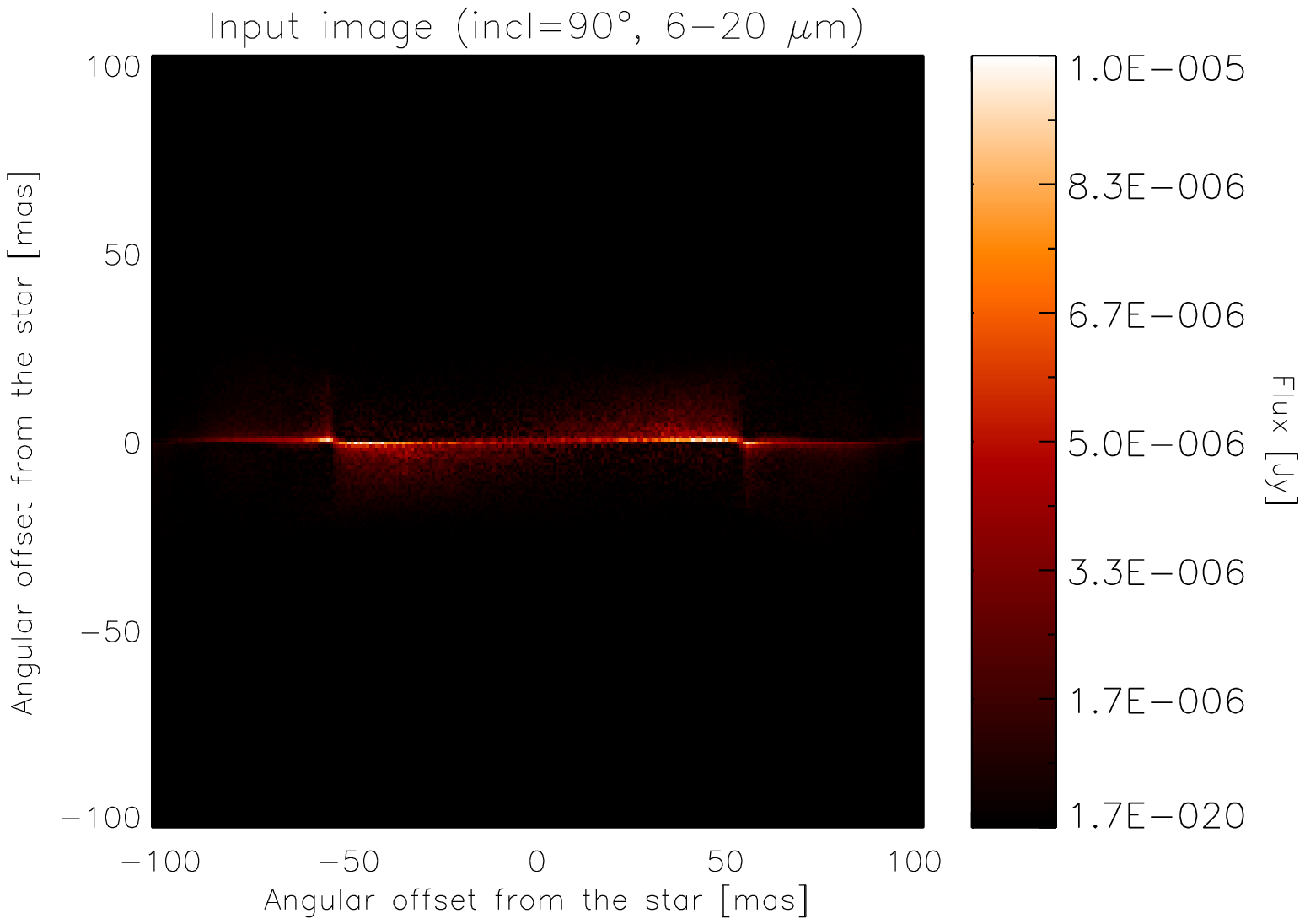}\hspace{8mm}
\caption{\emph{Upper:} Thermal flux (6-20\,$\mu$m) produced by a 10-zodi exozodiacal dust cloud around a G2V star for four different disc inclinations (0$^{\circ}$, 30$^{\circ}$, 60$^{\circ}$ and 90$^{\circ}$) and assuming a Dohnanyi distribution of particle sizes \citep{Dohnanyi:1969}. The images have been simulated assuming a Earth-mass planet located at 1\,AU on the x-axis (90 degrees clockwise from vertical,  \citealt{Stark:2008}). \emph{Lower:} Corresponding asymmetric brightness distributions obtained by subtracting to each pixel its centrally symmetric counterpart.}\label{fig:disc_images}
\end{center}
\end{figure*}

The previous results have been obtained assuming that the exozodiacal dust emission is
centrally symmetric around the target star so that it is suppressed by phase chopping
(and therefore only contributes to shot noise). However, exozodiacal discs are
likely to show resonant structures or an offset with respect to the central star
due to the the gravitational influence of embedded planets. These resonant structures
have been predicted by theoretical studies \citep{Roques:1994,Liou:1999,Ozernoy:2000}, and similar structures have been observed
in few cases around nearby main-sequence stars (e.g.,\ \citealt{Wilner:2002,Greaves:2005,Kalas:2005,Schneider:2009}).
The most well-studied example of an asymmetric disc is the solar zodiacal cloud, which exhibits several structures
interpreted as the dynamical signature of planets \citep{Dermott:1985,Dermott:1994,Reach:1995}. This trend suggests
that exozodiacal clouds may be full of rings, clumps, and other asymmetries induced by the presence of embedded planets.
These asymmetric structures around the target star are not perfectly canceled by the phase chopping process and part of the exozodi
signal can then mimic the planetary signal. If the demodulated contribution from the exozodiacal disc is significantly higher
than that of the planet, it would be difficult to isolate the planetary signal, whatever the integration time.
As mentioned by \cite{Lay:2004}, asymmetric inhomogeneities at the 0.1\% level of the total exozodiacal flux
can be confused with a planetary signal. The problem becomes even more serious as the dust density increases:
a 10-zodi exozodiacal disc must be smooth at the 0.01\% level, in the region to be searched for planets.
Considering asymmetric exozodiacal discs, we derive in this section the tolerable dust density in order to ensure the detection
of Earth-like planets with \darwin/TPF.

\subsection{Methodology}

At the output of the interferometer, the total detected photon rate (excluding
stray light) can be written as \citep{Lay:2004}:

\begin{equation}
N=\int_\theta\int_\alpha B_{{\rm
sky}}(\textbf{s})R(\textbf{s})P(\textbf{s})\,\theta d\theta
d\alpha\, , \label{eq:N}
\end{equation}
where $B_{{\rm sky}}$ is the brightness distribution on the sky
for a spectral channel centered on wavelength $\lambda$ and a
bandwidth $\Delta\lambda\ll\lambda$ (units in photons/s/{\rm
m$^2$}),~{\bf s} is a unit vector whose direction represents
position on the sky, P({\bf s}) is a field-of-view taper function
resulting from the size of a collecting aperture and from the
response of the single-mode spatial filter, and $R({\bf s})$ is
the intensity response of the interferometer on the sky (excluding the taper),
the so-called transmission map. Expression~\ref{eq:N} can be written as a function of the double Fourier
transform of the sky brightness, including the field-of-view taper, which are
denoted by a horizontal bar:

\begin{align}
N=\sum_j\sum_kA_jA_k\left[\cos(\phi_j-\phi_k)\right.&\overline{B}_{\rm sky,jk,sym}\nonumber\\
-&\left.\sin(\phi_j-\phi_k)\overline{B}_{\rm sky,jk,asym}\right]\label{eq:N3}\, ,
\end{align}
where $B_{{\rm sky,sym}}$ (resp.~$B_{{\rm sky,asym}}$) is the symmetric
(resp.~asymmetric) part of the sky brightness distribution and $\phi_j$ the phase response of the telescope $j$. The total
photon rate is therefore a sum over all possible pairs of collectors.
The baselines with a phase difference that is an integer multiple of $\pi$
(0, $\pm\pi$, $\pm2\pi$) couple entirely to the symmetric brightness
distribution (star, local zodiacal cloud, exozodiacal disc). Any asymmetric
brightness distribution couples entirely to baselines with a phase difference
that is an odd multiple of $\pi/2$ (e.g.,\, planet, clump). From Eq.~\ref{eq:N3},
the detected photon rate from the exozodiacal disc is simply given by:
\begin{align}
N_{\rm EZ}=\sum_j\sum_kA_jA_k\left[\cos(\phi_j-\phi_k)\right.&\overline{B}_{{\rm EZ,jk,sym}}\nonumber\\
-&\left.\sin(\phi_j-\phi_k)\overline{B}_{\rm EZ,jk,asym}\right]\label{eq:Nez}\, ,
\end{align}
The demodulated signal from the exozodiacal disc can then be obtained (after combination of the two chop states):
\begin{equation}
O_{\rm EZ}=\frac{1}{T}\int_0^T\left(N^L_{\rm EZ}-N^R_{\rm EZ}\right)\eta dt \label{eq:Nez2}\, ,
\end{equation}
where $T$ is the integration time, N$^L_{\rm EZ}$ (resp.\ N$^R_{\rm EZ}$)
the contribution of the exozodiacal disc at the output of the left (resp.\ right) chop state and $\eta$ the demodulation template
function. This template function is used to extract the planetary signal
by cross-correlation \citep{Angel:1997}. It represents the time series (normalized to have a
rms value of one) that would be obtained, should a planet be present
at a given position. Taking into account that the phase response of the interferometer is such that
$\phi_i^L=-\phi_i^R$ with the $\phi_i^L$ equal to ($\pi$/2, 0, $\pi$, 3$\pi$/2), Eq.~\ref{eq:Nez} and \ref{eq:Nez2}
give:

\begin{equation}
O_{\rm EZ}=\frac{2}{T}\sum_j\sum_kA_jA_k\sin(\phi_j-\phi_k)\int_0^T\overline{B}_{\rm EZ,jk,asym}\eta dt \label{eq:a_disc}\, ,
\end{equation}
which shows that only asymmetric components of the brightness distribution contribute to the demodulated signal.
More specifically, only the baselines with a ``fractional-$\pi$" phase difference can produce a demodulated signal.
Such baselines are mandatory in order to generate odd harmonics of the rotation frequency.
A solution to mitigate the influence of the asymmetric structures in the disc is to have long imaging (``fractional-$\pi$")
baselines that resolve out the more spatially extended emission from the exozodi variations, leaving only the point like emission
from planets.

\subsection{Impact of clumps}

\begin{figure*}[!t]
\begin{center}
\includegraphics[width=4.5 cm]{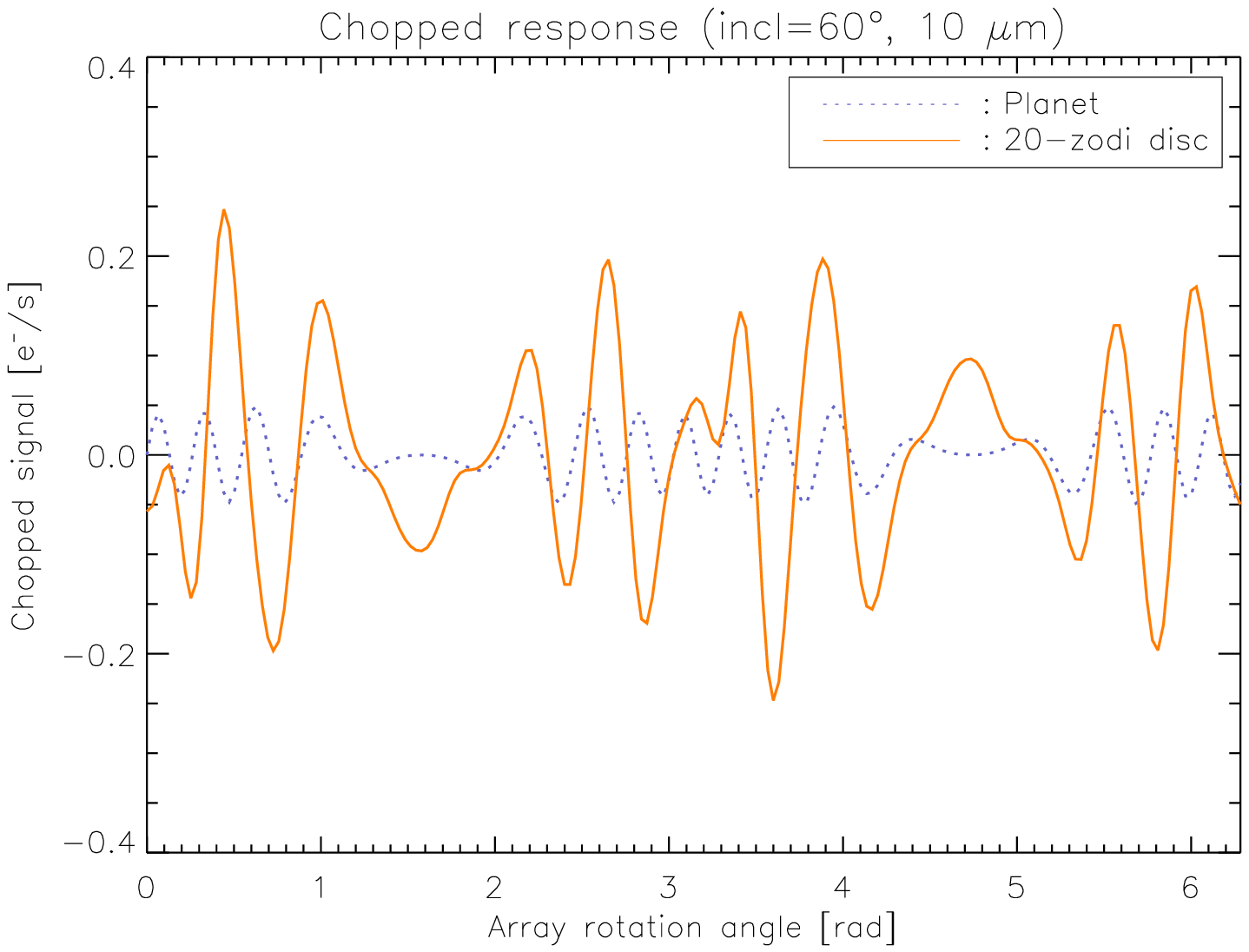}
\includegraphics[width=4.5 cm]{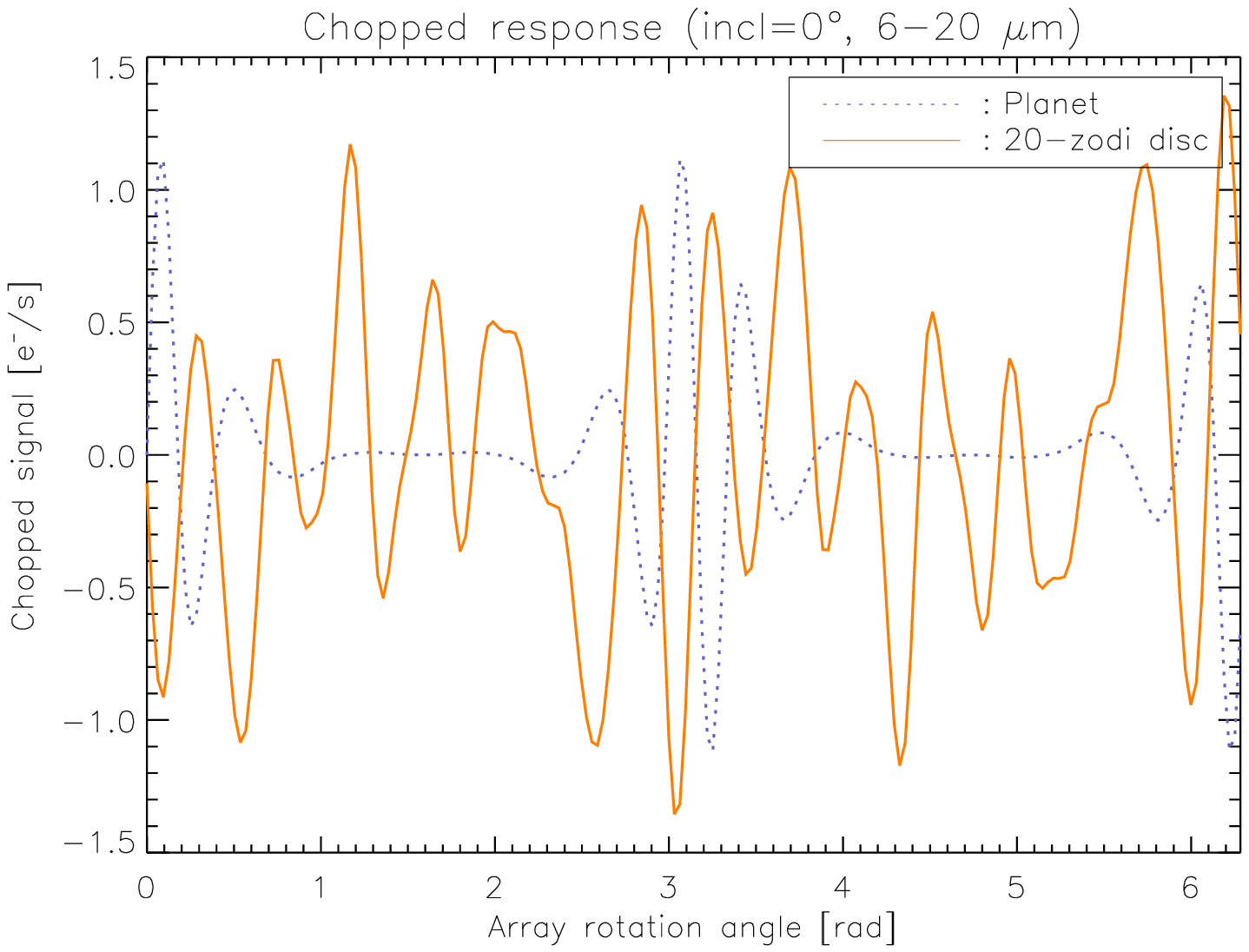}
\includegraphics[width=4.5 cm]{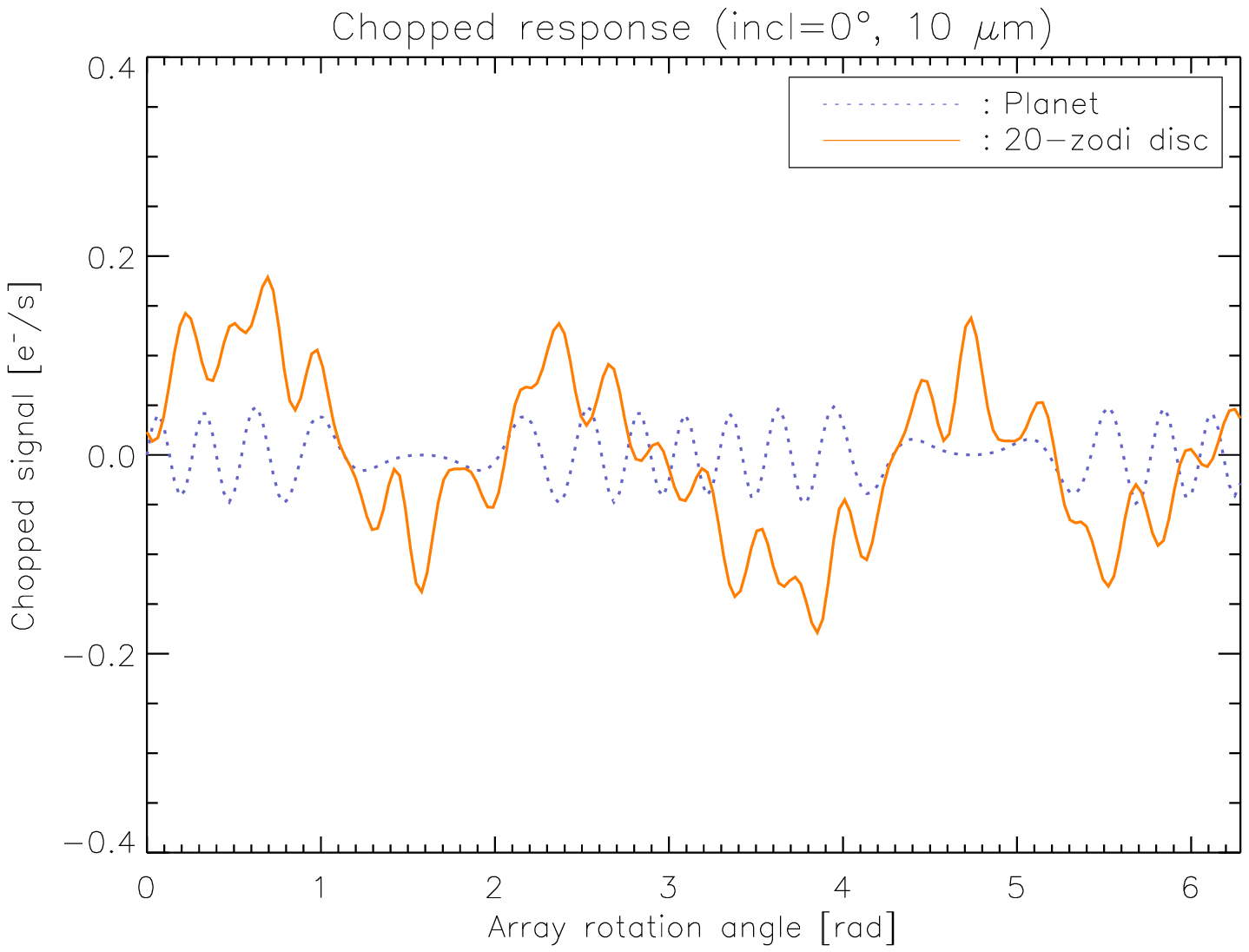}
\includegraphics[width=4.5 cm]{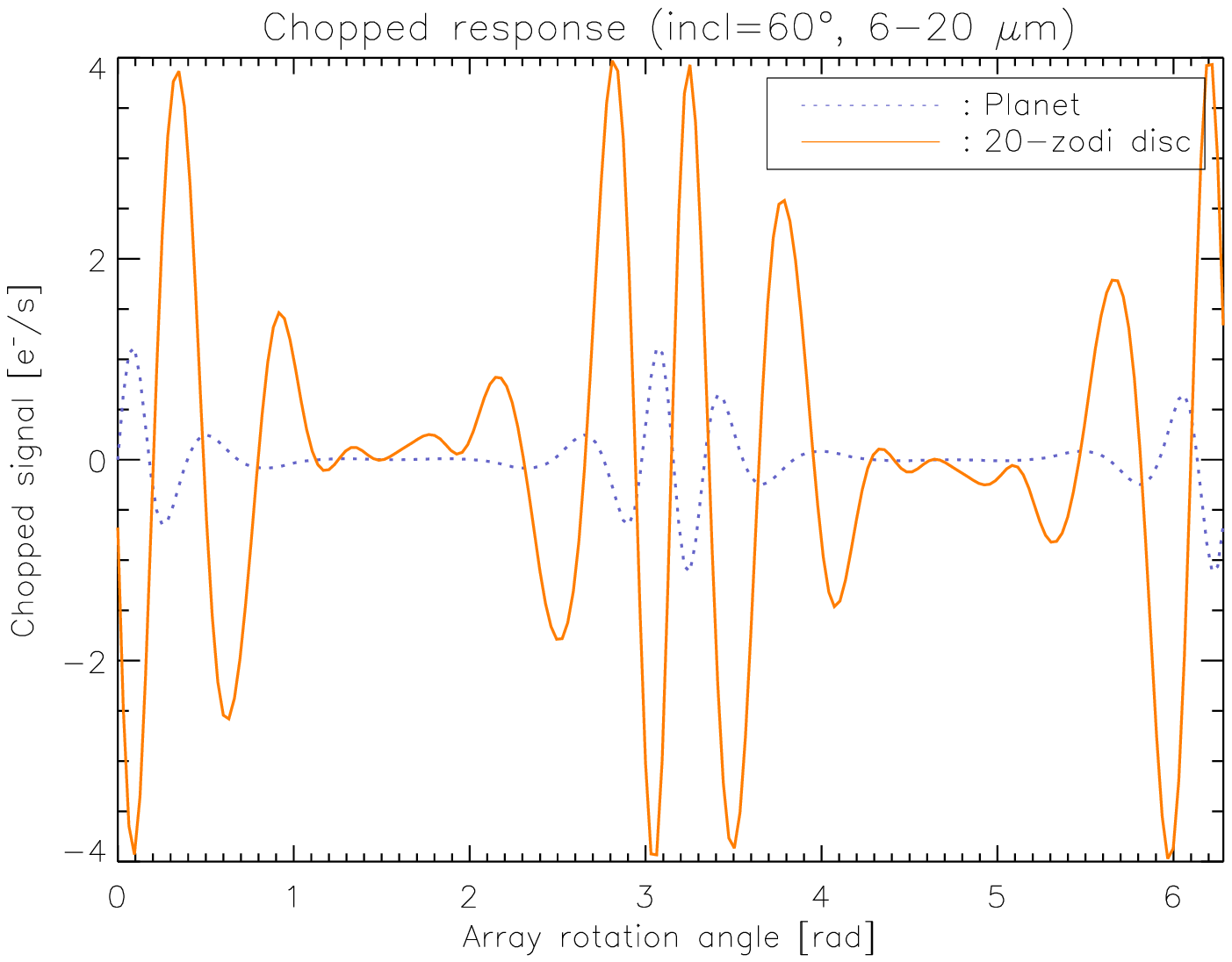}
\includegraphics[width=4.5 cm]{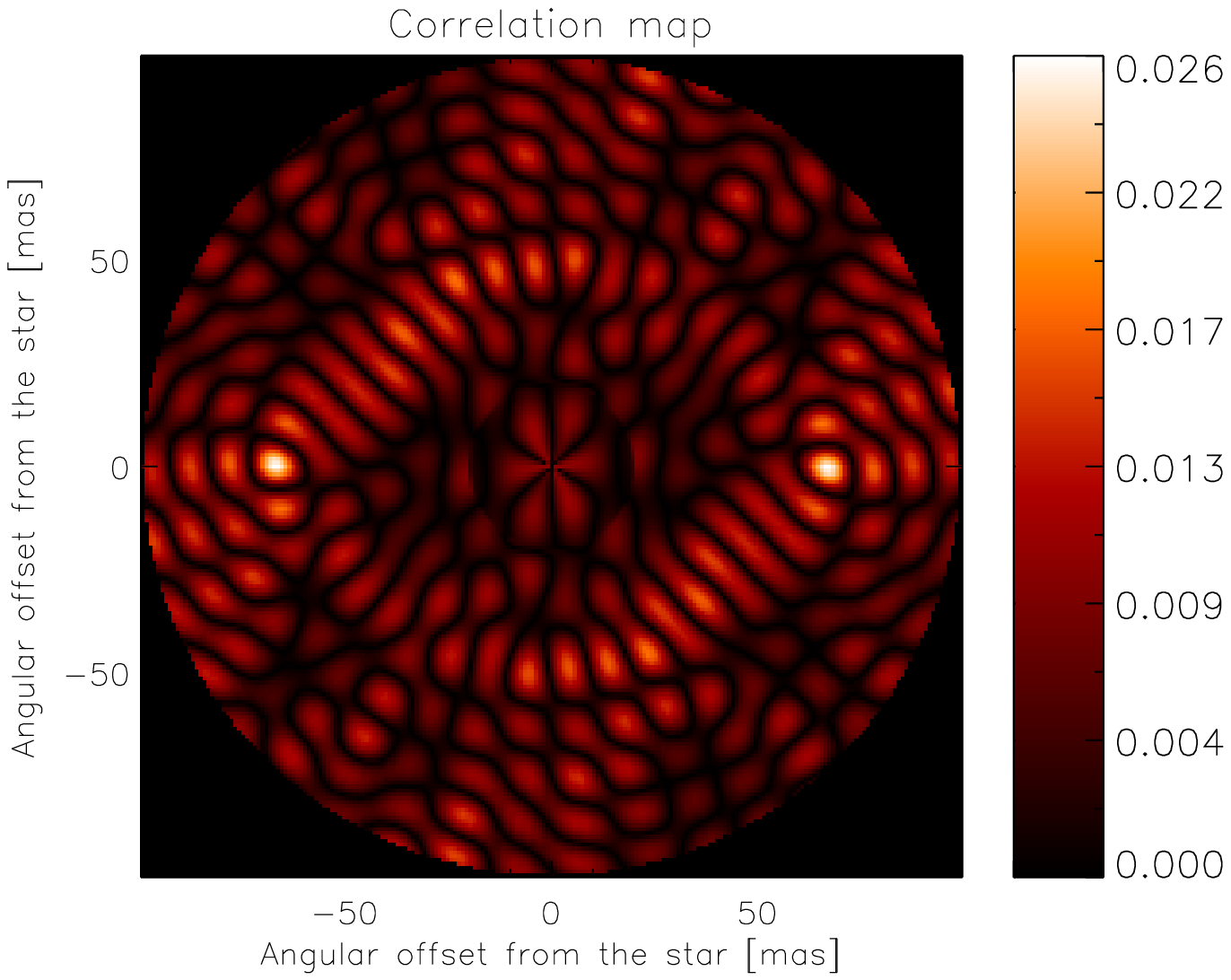}
\includegraphics[width=4.5 cm]{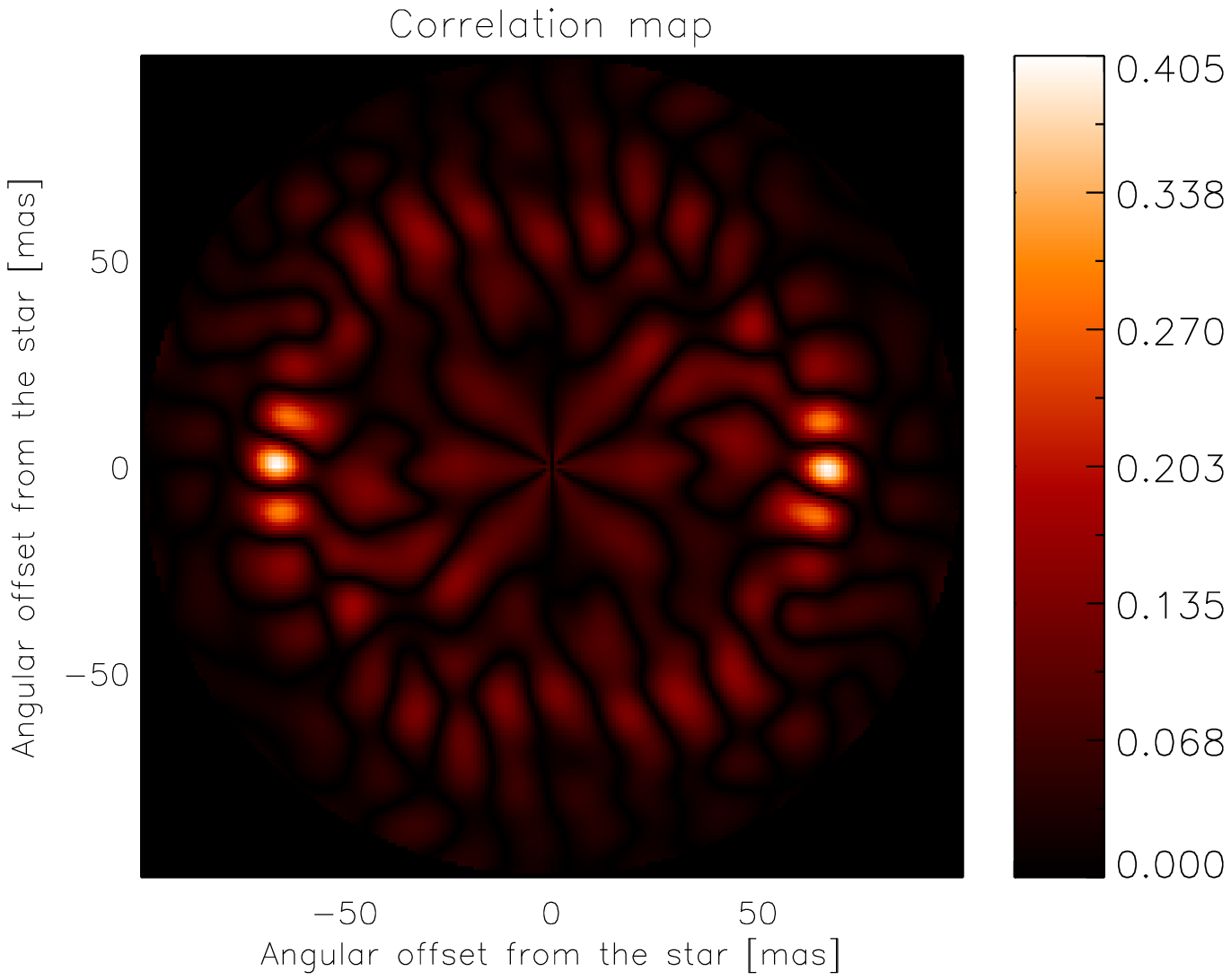}
\includegraphics[width=4.5 cm]{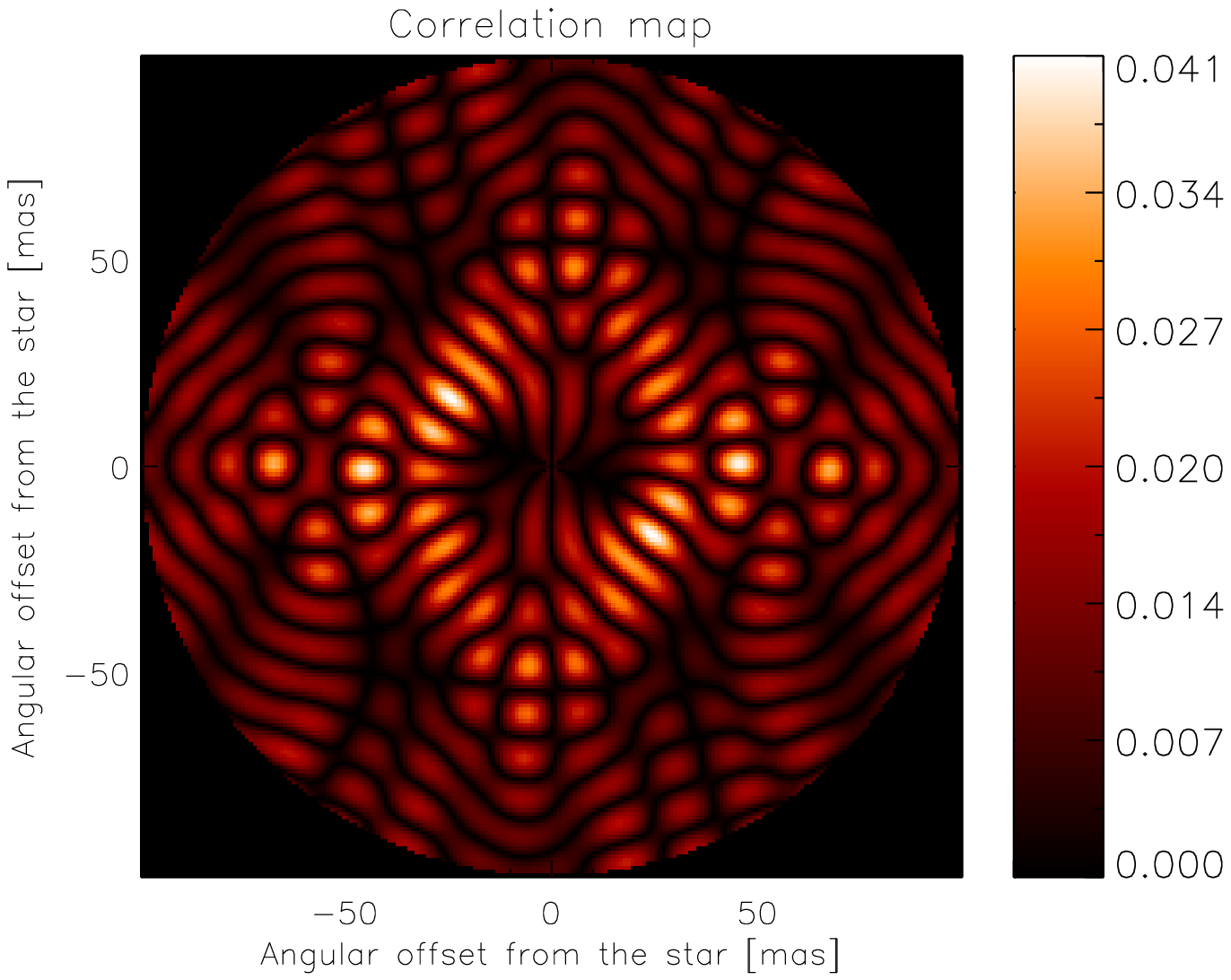}
\includegraphics[width=4.5 cm]{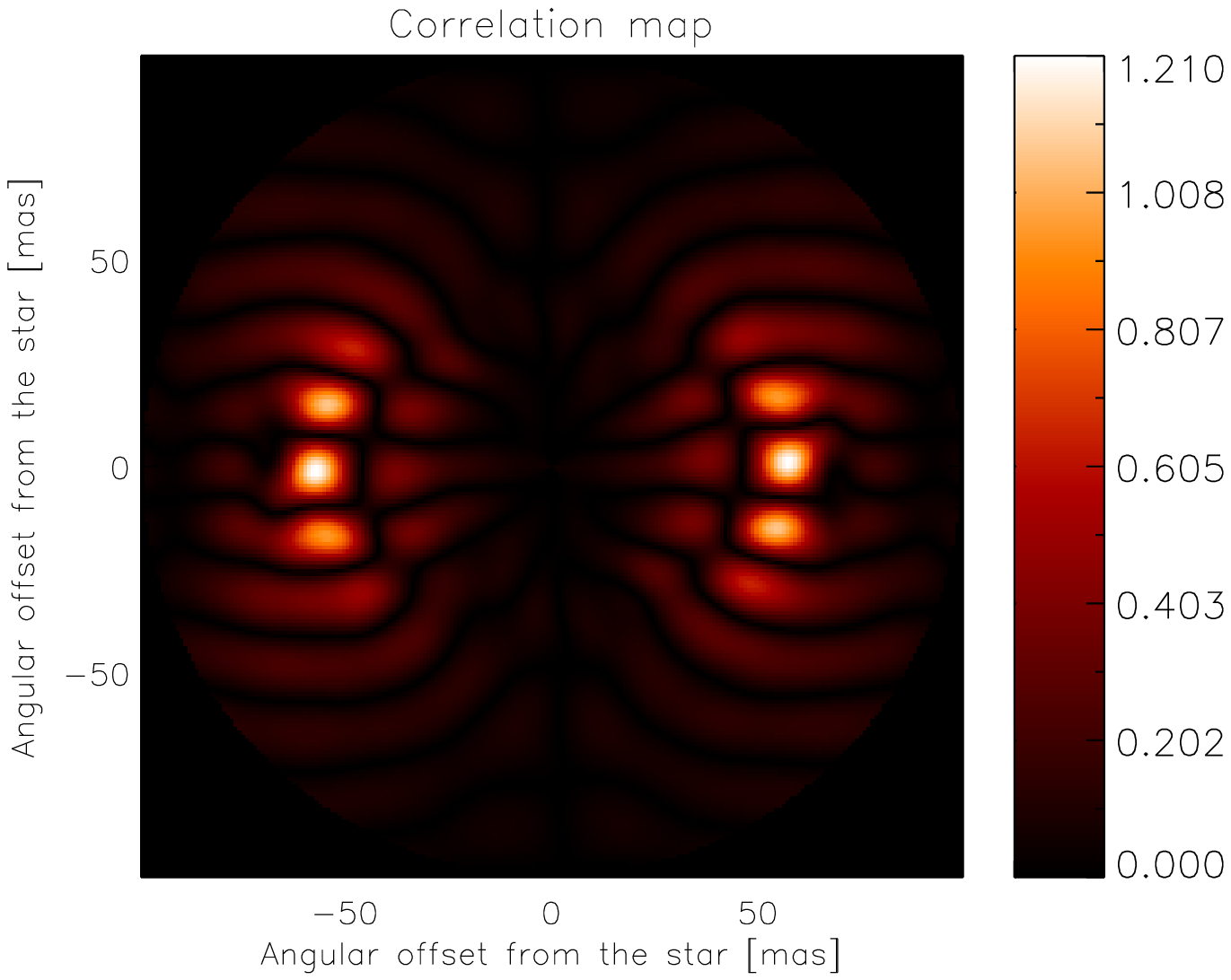}
\caption{\emph{Upper:} Chopped photon rate from an Earth-like planet and a 10-zodi asymmetric disc with respect to the rotation angle for two disc inclinations (0$^{\circ}$ and 60$^{\circ}$) and different wavelength ranges (at 10\,$\mu$m and in the full wavelength range). \emph{Lower:} Corresponding dirty map formed from the cross correlation of the measured signal with templates of the signal expected from a point source at each location on the sky.}\label{Fig:zodi_vs_ang}
\end{center}
\end{figure*}

The origin of some asymmetric clumpy structures, i.e. local density enhancements, in exozodiacal discs may be attributed to to the gravitational influence of planets on the small dust grains. After their release from parent bodies via collisions or outgassing, dust grains experience
different paths in the stellar system, depending on their effective size. Whereas the smallest
particles are ejected from the planetary systems by radiation pressure in a dynamical time, larger
particles slowly spiral inward due to Pointing-Robertson drag \citep{Robertson:1937}.
While spiraling toward their host star, dust particles may become
temporarily trapped in mean motion resonance with planets, extending
their lifetimes. This trapping locally enhances
the particle density, creating structures, originally described for the solar zodiacal
cloud as circumstellar rings, bands, and clumps (e.g., \citealt{Kelsall:1998}).

In order to address the impact of such structures on the performance of \darwin/TPF, we use the results of
\cite{Stark:2008}, who synthesized images of circumstellar discs with resonant rings structures due to
embedded terrestrial-mass planets. Among the studies that have examined the geometry of these resonant signatures (e.g., \citealt{Kuchner:2003,Moro-Martin:2005,Reche:2008,Stark:2008}),
these images are particularly convenient for our study since they include enough particles to overcome the limitations of previous simulations, which were often dominated by various sources of Poisson noise, and allow for quantitative study of the modeled ring structures.  In addition, these images are geared toward terrestrial-mass planets at a few AU from the star, whereas most other studies concern more massive planets located much farther from the star. We used the \cite{Stark:2008} models to produce thermal emission images of inclined discs (0$^\circ$, 30$^\circ$, 60$^\circ$ and 90$^\circ$) with resonant ring structures. We investigated disc models for a system with an Earth-mass planet on a circular orbit at 1\,AU around a G2V star located at 15\,pc and for a Dohnanyi distribution ranging in size from the blowout size up to 120\,$\mu$m \citep{Dohnanyi:1969}.
The thermal emission produced by such exozodiacal discs are given in the upper part of Fig.~\ref{fig:disc_images} in a wavelength range of 6-20\,$\mu$m.

The images given in Fig.~\ref{fig:disc_images} can be thought of as upper-limits to the brightness of structures due to an Earth-like planet. \cite{Stark:2008} ignored dust from parent bodies with large inclinations and eccentricities, such as comets, which would tend to wash out any resonant structure. Additionally, these models ignore the effects of collisions, which smooth out overdense regions of the disc and reduce azimuthal asymmetries \citep{Stark:2009}. In every simulation, the parent bodies were initially distributed from 3.5 to 4.5\,AU in an asteroid belt-like ring. The Earth-mass planet is oriented along the x-axis (located at 66\,mas on the x-axis) with a noticeable gap in the ring at its position. The models are truncated at half the semi-major axis of the planet, resulting in the inner holes in the images of Fig.~\ref{fig:disc_images}. In reality, the dust density distribution should continue inward to the dust sublimation radius in the absence of additional perturbers. This ``missing" inner disc should however not affect our results, since the inner disc would be centrally symmetric so that it does not contribute to the detected signal.

Introducing these images into \darwinsim, we compute the chopped photon rate from the exozodiacal disc as a function of the array rotation angle. This is represented in the upper part of Fig.~\ref{Fig:zodi_vs_ang} for two different disc inclinations (0$^\circ$ and 60$^\circ$) at 10\,$\mu$m (left figures) and for broadband detection (6-20\,$\mu$m, right figures). The density of the disc has been scaled up to 10 zodis and the chopped planetary signal represented for comparison (dashed curve). In all cases, it is dominated by the chopped signal from the exozodiacal disc (solid curve) and particularly for high disc inclinations. To disentangle the planetary signal from the disc signal, it is necessary to apply the cross-correlation method to build the so-called \textit{dirty map}. Applying cross correlation of the measured signal (disc + planet) with templates of the signal expected from a point source at each location on the sky (computed using Eq.~\ref{eq:a_disc}), the dirty maps represented in the lower part of Fig.~\ref{Fig:zodi_vs_ang} are obtained. This process transforms the rotationally modulated signal into a map of the sky by cross-correlation, which is equivalent to the Fourier transform used in standard synthesis imaging. The planet is located at 66\,mas on the x-axis and presents a demodulated signal of about 0.028\,e$^-$/s at 10\,$\mu$m and about 0.62\,$e^-$/s for broadband detection. The total demodulated signal at the planet position is however lower due to the negative contribution from the exozodiacal disc which presents a hole near the planet. The demodulation of the exozodiacal discs also produces main peaks which are maximum around 30-50\,mas from the host star (in agreement with the asymmetric brightness distribution in the initial images, see the lower part of Fig.~\ref{fig:disc_images}). Fortunately, the high angular resolution provided by the long imaging baseline is sufficient to spatially distinguish these components from the planetary signal and only the contribution from the hole around the planet significantly contributes to the noise level.

In order to ensure the planet detection, we adopt a criterion commonly used in AO imaging \citep{Macintosh:2003, Hinkley:2007,Serabyn:2009}. The noise level is taken to be the rms deviation of the pixel counts within an annulus of width equal to the size of the PSF at half maximum. Considering a detection threshold of 5, the results are given in Table~\ref{tab:detectivity} for different wavelengths and disc inclinations. The tolerable disc density ranges between about 1 and 15 zodis, depending on the disc inclination. The detection is particularly difficult for highly inclined discs for which the asymmetric components are more dominant and at long wavelengths where the planetary signal is weaker. Combining the spectral channels to obtain a broadband correlation map (see Fig.~\ref{Fig:zodi_vs_ang}) reduces the impact of sidelobes associated with each main peak but does not significantly improve the results. This is because the performance are limited by the main peak induced by the hole near the planet rather than by sidelobes. Note that this hole generally induces a response 2 to 4 times larger the rms deviation of the pixel counts within the annulus so that it might be marginally interpreted as a planet detection (a ``false positive''). However, it could also be seen as an indirect way to detect the presence of a planet within the hole since such compact structure is expected to be created by a planet.

In order to relax these stringent constraints on the exozodiacal dust density, more sophisticated techniques of data processing could be useful (e.g.,\,\citealt{Thiebaut:2006}). The capability of these techniques still needs to be investigated and is beyond the scope of this paper. The only secure way to ensure the detection and characterization of Earth-like planets with \darwin/TPF-like missions is to observe in advance the nearby main-sequence stars in order to remove from the target list  the stars with a too high inclined/bright exozodiacal disc. To achieve this goal, a space-based nulling interferometer such as FKSI (``Fourier-Kelvin Stellar Interferometer") would be ideal with a sensitivity sufficient to detect exozodiacal discs down to 1 zodi \citep{Defrere:2007c}.


\subsection{Impact of the disc offset}

\begin{figure}[!t]
\begin{center}
\includegraphics[width=8.5 cm]{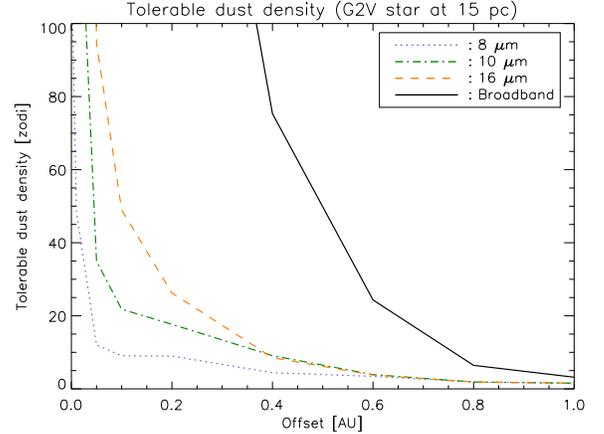}
\caption{Tolerable exozodiacal dust density with respect to the offset between the center of symmetry of the exozodiacal disc and the central star (a G2V star located at 15\,pc). The disc is assumed to be seen in face-on orientation.}\label{Fig:zodi_vs_off}
\end{center}
\end{figure}

An offset between the center of symmetry of a dust cloud and its host star is a natural consequence of the
gravitational interaction with planets. In the solar system, the center of the zodiacal cloud
is shifted by about 0.013\,AU from the Sun due mostly to Jupiter \citep{Landgraf:2001}. The offset
can be much larger, as shown in the case of the Fomalhaut system with an offset of 15\,AU \citep{Kalas:2005}.
Even when inhomogeneities such as clumps are not present, an offset cloud produces an asymmetric brightness distribution
such that a part of the exozodiacal disc signal survives the chopping process. Using the zodipic package\footnote{\textit{http;//asd.gsfc.nasa.gov/Marc.Kuchner/home.html}}, we produce
images of solar-like zodiacal discs with a given offset and use them to compute the demodulated signal at the
output of the interferometer. The results are presented in Fig.~\ref{Fig:zodi_vs_off}, showing the tolerable dust density with
respect to the disc offset for a G2V star located at 15\,pc and for different wavelengths. The disc is assumed to be seen face-on.

As the distance between the host star and the center of symmetry of the exozodiacal disc increases, the tolerable dust density to
detect an Earth-like planet located at 1\,AU  becomes more severe and reaches less than 5 zodis for an offset of 1\,AU. For individual
spectral channels, the tolerable dust density rapidly decreases to reach the value of 20 zodis at 0.05, 0.15 and 0.25 AU respectively
at 8\,$\mu$m, 10\,$\mu$m and 16\,$\mu$m. The results for broadband detection are much better with a tolerable dust density of 20 zodis
only for a disc presenting an offset larger than 0.6\,AU. Considering a tolerable exozodiacal dust density of 100 zodis, the offset
between the host star and the center of symmetry of the exozodiacal disc can be as high as 0.4 AU in order to ensure the planet detection.
For an offset similar to that of the solar zodiacal cloud (about 0.013\,AU), this tolerable dust density is even much higher (few thousand
zodis).

\begin{table}[!t]
    \centering
    \caption{Tolerable exozodiacal dust density for different disc inclinations and wavelengths. The detection threshold is taken to be 5 times the rms deviation of the pixel counts within an annulus of width equal to the size of the PSF at half maximum.}\label{tab:detectivity}
    \begin{tabular}{c c c c c}
      \hline
      \hline
      Disc incl. & 8\,$\mu$m & 10\,$\mu$m & 16\,$\mu$m & Wide \\
      \hline
      0$^\circ$ & 12.2 & 15.3 & 7.0 & 14.0 \\
      30$^\circ$ & 9.1 & 15.0 & 6.9 & 13.8 \\
      60$^\circ$ & 6.1 & 6.8 & 2.1 & 3.8 \\
      90$^\circ$ & 1.1 & 1.4 & 1.0 & 2.6 \\
      \hline
    \end{tabular}
\end{table}


%

\section{Conclusions}

Infrared nulling interferometry is the core technique of future
life-finding space missions such as ESA's \darwin{} and NASA's
Terrestrial Planet Finder (TPF). Observing in the infrared (6-20
$\mu$m), these missions will be able to characterise the
atmosphere of habitable extrasolar planets orbiting around nearby
main sequence stars. This ability to study distant planets
strongly depends on exozodiacal clouds around the stars, which can
hamper the planet detection. Considering the nominal mission architecture
with 2-m aperture telescopes, we show that centrally symmetric exozodiacal
dust discs about 100 times denser than the solar zodiacal cloud can be tolerated in
order to survey at least 150 targets during the mission lifetime. The actual number
of planet detections will then depend on the number of terrestrial planets in the habitable zone of
target systems.

The presence of asymmetric structures in exozodiacal discs (e.g.,\ clumps or offset) may be more
problematic. While the cloud brightness drives the integration time necessary to disentangle the
planetary photons from the background noise, the emission from inhomogeneities are not perfectly
subtracted by phase chopping so that a part of the disc signal can mimic the planet.
To address this issue, we consider modeled resonant structures produced by an Earth-like planet
and introduce the corresponding image into \darwinsim, the mission science simulator. Even for exozodiacal
discs a few times brighter than the solar zodiacal cloud, the contribution of these
asymmetric structures can be much larger than the planetary signal at the output of the interferometer.
Fortunately, the high angular resolution provided by the long imaging baseline of \darwin/TPF in the X-array configuration
is sufficient to spatially distinguish most of the extended exozodi emission from the planetary signal and
only the hole in the dust distribution near the planet significantly contributes to the noise level.
Considering the full wavelength range of \darwin/TPF, we show that the tolerable dust density is about 15 times
the solar zodiacal density for face-on systems and decreases with the disc inclination. In practice, this constraint
might be relaxed since we examined a resonant ring model that does not include dust from highly eccentric or inclined
parent bodies, the effects of grain-grain collisions, or perturbations by additional planets, all of which can reduce
the contrast of the resonant ring and improve the tolerance to the exozodiacal dust density. 


These results show that asymmetric structures in exozodiacal discs around nearby main sequence stars are one of the main noise sources
for future exo-Earth characterization missions. A first solution to get around this issue is to have a long imaging baseline
architecture which resolves out the more spatially extended emission of the exozodiacal cloud from
the point-like emission of planets. The stretched X-array configuration is particularly convenient in that respect.
The second solution is to observe in advance the nearby main sequence stars and remove from the \darwin/TPF target
list those presenting a too high dust density or disc inclination. The FKSI nulling interferometer would be ideal in that respect with
the possibility to detect exozodiacal discs down to the density of the solar zodiacal cloud. Ground-based nulling
instruments like LBTI and ALADDIN would also be particularly valuable.

\appendix
\section{Deriving instability noise constraints} \label{app:instab}

\begin{figure}[!t]
\begin{center}
\includegraphics[width=8.5 cm]{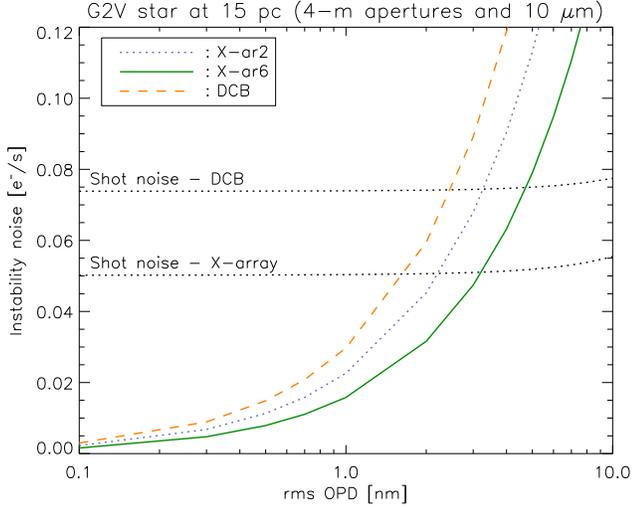}
\caption{Instability noise with respect to the rms OPD errors for different array architectures assuming 1/f-type PSDs and a rms amplitude mismatch of 0.1\% (defined on the 0-10$^4$\,Hz frequency range). The level of shot noise is represented by dotted curves for each configuration. The figure has been plotted for 4-m aperture telescopes operating at 10\,$\mu$m and a G2V star located at 15\,pc (surrounded by an exozodiacal cloud of 1 zodi).}\label{Fig:in_vs_sn}
\end{center}
\end{figure}

We address here the instability noise and derive the constraints on the instrument stability
required to detect an Earth-like planet orbiting at 1\,AU around a G2V star located at 15\,pc.
The analysis follows the analytical method of \cite{Lay:2004} which was originally applied
to the DCB configuration at 10\,$\mu$m. The goal is to extend the study to the X-array architecture
and to short wavelengths where instability noise is the most dominant. We define the limiting rms OPD and amplitude errors such that instability noise is dominated by a factor 5 by shot noise over a single rotation of 50000\,s. Assuming that instability noise is totally uncorrelated with the rotation angle, this factor stays unchanged over multiple rotations. In practice, instability noise can be correlated with the rotation angle
due to perturbations such as solar heating effects but it should be possible to remove them by measuring and correcting the amplitudes and phases at intervals during the rotation.

\begin{figure}[!t]
\begin{center}
\includegraphics[width=8.5 cm]{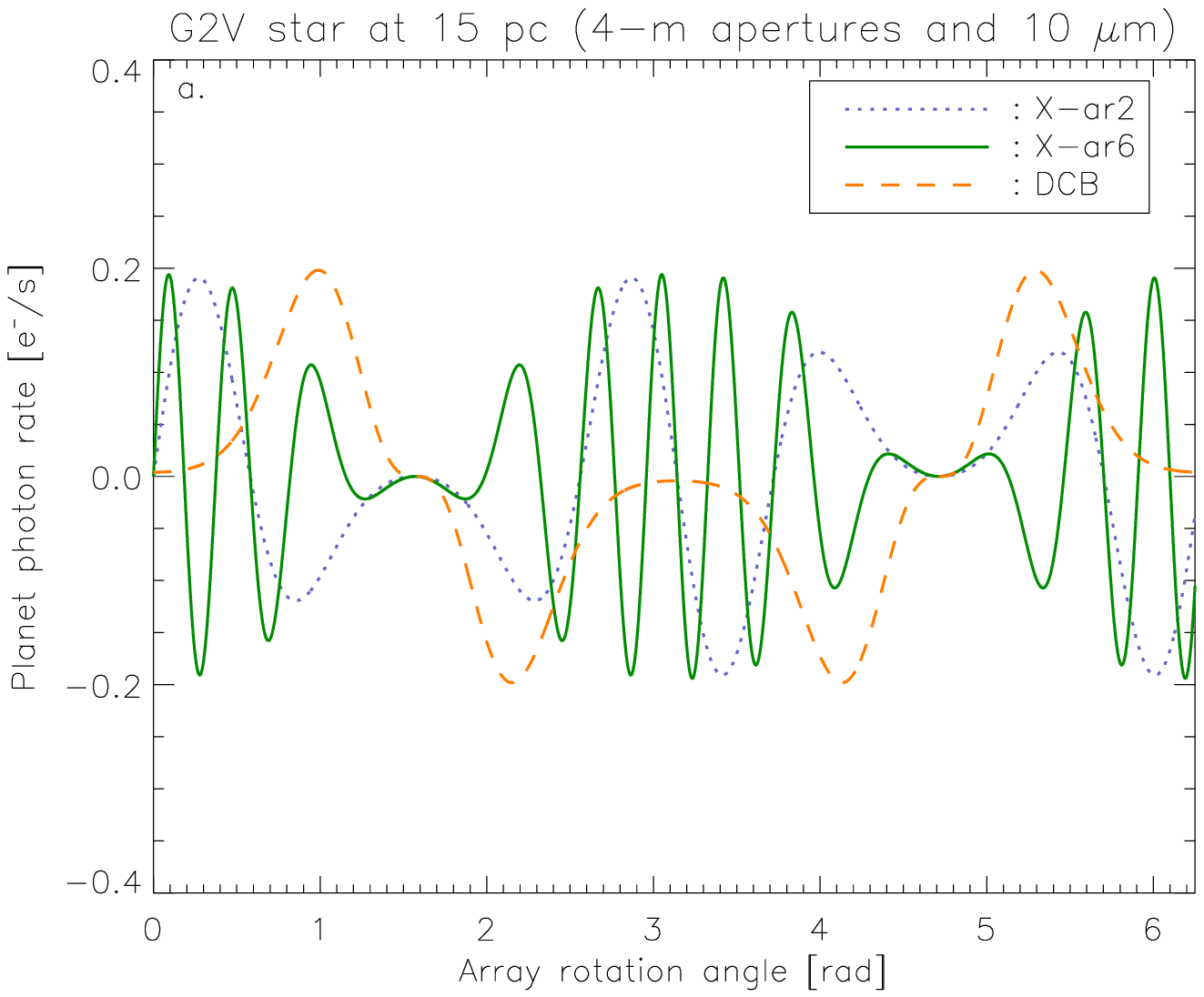}
\includegraphics[width=8.5 cm]{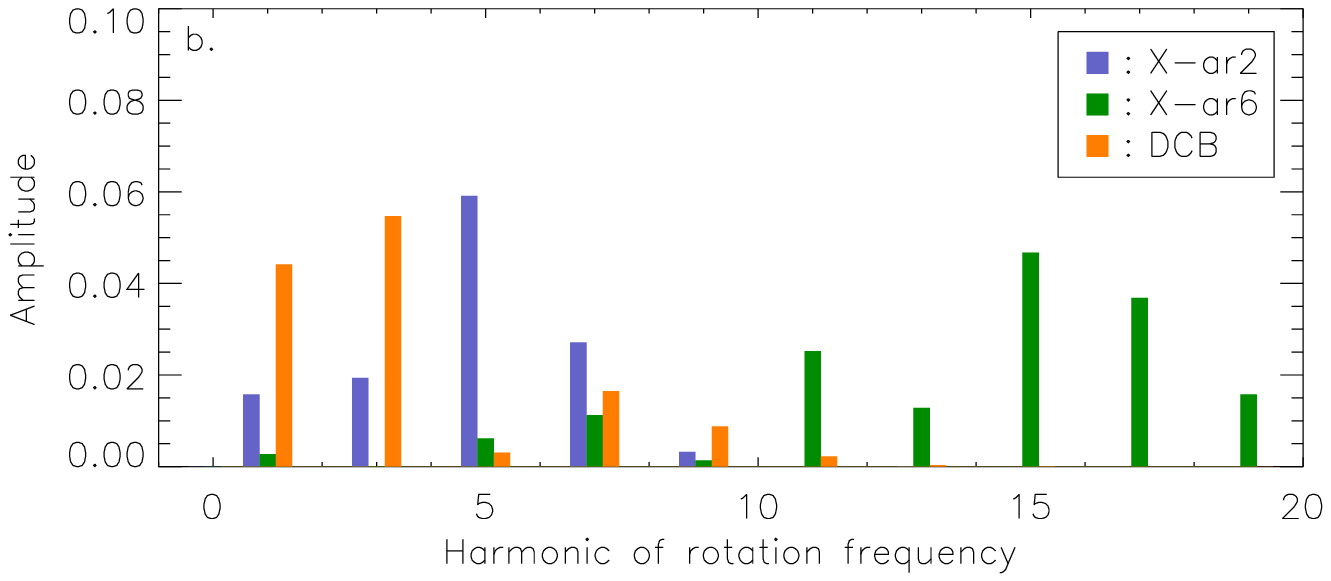}
\caption{\emph{Upper:} Chopped planet detected photon rate as a function of array rotation angle for the different architectures. The planet is assumed to be located at 47 mas from a G2V star located at 15\,pc. \emph{Lower:} Corresponding Fourier amplitudes. Only odd harmonics are present because of phase chopping.}
\label{Fig:eta}
\end{center}
\end{figure}

For sake of comparison with \cite{Lay:2004}, we consider first 4-m aperture telescopes operating at 10\,$\mu$m and extend the results to 2-m aperture telescopes operating at 7\,$\mu$m in Table~\ref{tab:instab_noi}. Assuming a spectral resolution of 20, Fig.~\ref{Fig:in_vs_sn} shows
instability noise with respect to the rms OPD error for three different configurations: the DCB as
defined in \cite{Lay:2004}, the X-array with a 1:2 aspect ratio (X-ar2) and the X-array with a 1:6 aspect ratio (X-ar6).
These curves have been computed assuming 1/f-type PSDs for amplitude and OPD errors with rms values defined on a frequency range from 1/t$_{\rm rot}$ to 10$^4$\,Hz, where t$_{\rm rot}$ is the rotation period, and a G2V star located at 15\,pc. Shot noise is represented by dotted curves. It is higher for the DCB due to geometrical leakage (the nulling baseline is twice larger than for the X-array) and presents a slight increase for large rms OPD errors due to instrumental leakage. Instability noise is also higher for the DCB than for the X-array. This is because the planetary signal is mostly modulated at lower frequencies where the instability noise is higher for 1/f-type noises. This is illustrated in Fig.~\ref{Fig:eta} showing the chopped planet detection rate with respect to the rotation angle of the array (upper figure) and the corresponding Fourier amplitudes (lower figure).

Fig.~\ref{Fig:in_vs_sn} shows that shot noise dominates instability noise by a factor 5 for rms OPD errors of about 0.5\,nm, 0.5\,nm and 0.6\,nm respectively for the DCB, X-ar2 and X-ar6. The slight discrepancy that can be mentioned with \cite{Lay:2004} is due to two factors. In addition to the instrument throughput of 10\% used in \cite{Lay:2004}, our study accounts for the coupling efficiency (about 72\% for the on-axis light) and for the quantum efficiency of the detectors (70\%). We also combine the two chop states whereas the results of \cite{Lay:2004} are given for only one.
Although instability noise is higher for the DCB configuration than for the X-array, the limiting rms opd error is of the same order due to the higher shot noise. Because instability noise is directly proportional to the stellar flux, the constraints are even more stringent at 7-$\mu$m where the star is brighter and the level of shot noise basically the same than at 10\,$\mu$m (see Table~\ref{tab:instab_noi}). Considering an rms amplitude mismatch of 0.1\%, the OPD has to be controlled to a level of 0.3\,nm rms for the three considered configurations. These constraints are slightly relaxed for 2-m aperture telescopes because shot noise is relatively more dominant (shot noise is proportional to the squared root of the stellar flux while instability is directly proportional to the stellar flux). For the results at 7\,$\mu$m, we use the same baseline length of 20\,m with the same rotation period of 50000\,s. These requirements are very stringent and are only marginally compliant with state-of-the-art active control, so that potential ways to mitigate the harmful effect of instability noise have been investigated.

\begin{table*}[!t]
\centering
\caption{Limiting rms OPD values computed for a G2V star located at 15\,pc such that shot noise dominates instability noise by a factor 5. The results are given at 7\,$\mu$m and 10\,$\mu$m, assuming a spectral resolution of 20, for different rms amplitude mismatches and shapes of the instability noise power spectrum. Three configurations are shown, the X-array with an aspect ratio of 2 (X-ar2), the X-array with an aspect ratio of 6 (X-ar6) and the Dual Chopped Bracewell (DCB, configuration used in \citealt{Lay:2004}).}
\setlength{\tabcolsep}{2pt}
\begin{tabular}{l c c}
  \hline \hline
  $\sigma_{\rm OPD}$ [nm] & 2-m apertures & 4-m apertures \\
  \begin{tabular}{l l}
                         &                  \\
                         & $\sigma_{\rm amp}$ \\
  \hline
  White noise$^a$ & 0.1\%  \\
                  & 0.5\%   \\
                  & 1.0\%   \\
  \hline
  1/f noise$^b$   & 0.01\%  \\
                  & 0.05\%   \\
                        & 0.10\%   \\
  \end{tabular}
  &
  \begin{tabular}{c c c c c c}
                              & $\lambda$=7\,$\mu$m & & & $\lambda$=10\,$\mu$m & \\
                          DCB & X-ar2 & X-ar6 & DCB & X-ar2 & X-ar6 \\
  \hline
                          46 & 17 & 17 & 97 & 95 & 95  \\
                          4.3 & 2.6 & 2.6 & 8.7 & 7.2 & 7.2 \\
                          2.3 & 1.5 & 1.5 & 4.5 & 3.8 & 3.8 \\
  \hline
                          0.7 & 2.6 & 4.2 & 2.8 & 13  & 19   \\
                          0.7 & 0.9 & 1.5 & 2.0 & 2.9 & 4.1  \\
                          0.5 & 0.5 & 0.7 & 1.5 & 1.6 & 2.0  \\
  \end{tabular}
  &
  \begin{tabular}{c c c c c c}
                              & $\lambda$=7\,$\mu$m & & & $\lambda$=10\,$\mu$m & \\
                          DCB & X-ar2 & X-ar6 & DCB & X-ar2 & X-ar6 \\
  \hline
                          10  & 5.5 & 5.5 & 19 & 14 & 14  \\
                          2.1 & 1.2 & 1.2 & 3.3 & 2.4 & 2.4  \\
                          1.1 & 0.6 & 0.6 & 1.8 & 1.4 & 1.4  \\
  \hline
                          0.4 & 1.1 & 1.9 & 1.1 & 3.6 & 5.6  \\
                          0.3 & 0.4 & 0.6 & 0.8 & 0.9 & 1.4  \\
                          0.3 & 0.3 & 0.3 & 0.5 & 0.5 & 0.6  \\
  \end{tabular}\\
  \hline
  \end{tabular}
  \begin{flushleft}
  {\small $^a$ rms values is defined on the [0-10\,Hz] wavelength range}\\
  {\small $^b$ rms values is defined on the [1/$t_{\rm rot}$-10$^4$\,Hz] wavelength range}
  \end{flushleft}
  \label{tab:instab_noi}
\end{table*}

\begin{itemize}
\item A first solution, proposed by \cite{Lay:2006}, consists in
stretching the array and applying a low-order polynomial fit to
the instability noise (as a function of wavelength). By stretching
the array, i.e.\ increasing the imaging baseline of the X-array,
the interference pattern orthogonal to the nulling pattern
shrinks. As the modulation map scales with wavelength, the
planetary signal transmitted by the interferometer will then be a
rapidly oscillation function of wavelength. On the other hand,
instability noise is shown to be a low-order polynomial of the
optical frequency (1/$\lambda$). Therefore, by removing a
low-order polynomial fit to the detected signal as a function of
wavelength, the instability noise contribution is efficiently
subtracted while preserving most of the planetary signal. This
operation can actually be performed directly in the
cross-correlation, by using a modified planet template where the
polynomial components have been removed. Because this method
strongly relies on the separation of the nulling and imaging
baselines, it can only be efficiently applied with the X-array
architecture. \item Another solution, based on the coherence
properties of starlight, has been proposed by \cite{Lane:2006} to
separate the contributions from the planet and the instrumental
leakage. The idea is to mix the electric fields of the leakage
with that of a separate reference beam, also from the star, in
order to form fringes (as long as the relative path delays are
maintained within the coherence length). The light from the
planet, being not coherent with the starlight, will not form
fringes. Using as reference beam the bright output of a pair-wise
nulling beam combiner, the amplitude and phase mismatches in the
input beams can be extracted from the fringe pattern, allowing the
reconstruction of the the stellar leakage. \item A third solution
has recently been proposed and tested at IAS \citep{Gabor:2008}.
The principle is to successively apply two ($<$ $\lambda$/2)
opposite OPD offsets to one of the beams in order to derive the
actual position of the minimum in the transmission map. In this
way, one feeds back the actual OPD errors to the delay line and
prevents drifts from appearing. The same principle can be used to
avoid amplitude drifts, by either blocking all but one beam to
measure its actual amplitude or by modulating its amplitude as in
the case of the OPD. The frequency at which this process is
carried out depends on the input power spectra of the noise
sources. It has been demonstrated experimentally in the lab that
this process efficiently suppresses the 1/f-type noise generally
present in the stellar contribution at the output of a Bracewell
interferometer. This technique can theoretically be applied to any
nulling configuration, but its efficiency decreases as the number
of beams increases.
\end{itemize}

Others possible approaches (e.g.\ application of closure phase
techniques, new chopping processes, exploiting correlations
between measurements taken at different wavelength bins) have been
suggested but still need to be investigated. However, these very stringent requirements have been derived with the assumption that the instability noise present an 1/f-type power spectrum. In fact, the shape of the instability noise spectrum strongly depends on the input spectrum of noise fluctuations. Pure 1/f noise might in fact turn out to be a very pessimistic assumption. Given that the \darwin/TPF system is
designed to have three levels of control loops that manage the OPD, and two control loop levels for tip/tilt, the resulting
residual phase and intensity fluctuations that are responsible for the instability noise are most likely to have a flat input
spectrum. This has a strong influence on the magnitude of the phase-amplitude cross terms, which are the main contributor to
instability noise. For instance, the tolerable rms OPD errors for a white spectrum defined on the [0-10\,Hz] frequency range are
significantly relaxed as indicated in Table~\ref{tab:instab_noi}. In particular at low frequencies, where the planetary signal is modulated and the instability noise arises, the 1/f spectrum diverges, while the flat spectrum increases linearly. In practice, the situation might even be better, as
predictive or Kalman filtering may be used to remove low-frequency power. A complication, in particular for the micro Hz domain, is
that zero point drifts of the control loops, e.g.\ related to electronic drifts in the sensors, remain undetected, and thus
uncorrected. They will result in a low-frequency 1/f component. A simple solution is to switch the incoming beams with respect to
the control loops. The reason why this could work is that the extremely high precision required for \darwin/TPF is only on relative
quantities, namely, the phase and amplitude differences between any two beams. Absolute offsets of phase and amplitude, that apply
to all three beams simultaneously, do not contribute to the instability noise, since they do not affect the null. The beams
must be switched with respect to the control inputs with a time constant over which the drift can be considered constant. In
summary, even though at present a post-processing method has not been identified, it is highly likely that clever and careful engineering may already
reduce instability noise to harmless levels.


\begin{acknowledgements}
The authors are grateful to Lisa Kaltenegger (Harvard-Smithsonian
Center for Astrophysics) for providing the updated \darwin{}
catalogue, Jean Surdej (IAGL), Arnaud Magette (IAGL), Dimitri Mawet (NASA/JPL), Peter Lawson (NASA/JPL), Oliver Lay (NASA/JPL),
Pierre Riaud (IAGL) and Virginie Chantry (IAGL). This research was supported by the International Space
Science Institute (ISSI) in Bern, Switzerland (``Exozodiacal Dust discs and Darwin" working group, \textit{http://www.issibern.ch/teams/exodust/}). DD and CH acknowledge the support of the Belgian National Science Foundation (``FRIA"). OA acknowledges the support from a F.R.S.-FNRS Postdoctoral Fellowship. DD and OA acknowledge support from the Communaut\'e fran\c{c}aise de Belgique - Actions de recherche concert\'ees - Acad\'emie universitaire Wallonie-Europe. 
\end{acknowledgements}

\bibliography{E:/PhD/biblio}

\begin{thebibliography}{71}
\expandafter\ifx\csname natexlab\endcsname\relax\def\natexlab#1{#1}\fi

\bibitem[{{Absil}(2001)}]{Absil:2001}
{Absil}, O. 2001, Master's thesis, Li\`ege University, Li\`ege, Belgium,
  \emph{http://www.aeos.ulg.ac.be/}

\bibitem[{{Absil}(2006)}]{Absil:thesis}
{Absil}, O. 2006, PhD thesis, Li\`ege University, Li\`ege, Belgium

\bibitem[{{Absil} {et~al.}(2008{\natexlab{a}}){Absil}, {Defr{\`e}re},
  {Coud{\'e} du Foresto}, {Di Folco}, {den Hartog}, \& {Augereau}}]{Absil:2008}
{Absil}, O., {Defr{\`e}re}, D., {Coud{\'e} du Foresto}, V., {et~al.}
  2008{\natexlab{a}}, in Proc. SPIE, Vol. 7013

\bibitem[{{Absil} {et~al.}(2006){Absil}, {di Folco}, {M{\'e}rand}, {Augereau},
  {Coud{\'e} Du Foresto}, {Aufdenberg}, {Kervella}, {Ridgway}, {Berger}, {Ten
  Brummelaar}, {Sturmann}, {Sturmann}, {Turner}, \& {McAlister}}]{Absil:2006b}
{Absil}, O., {di Folco}, E., {M{\'e}rand}, A., {et~al.} 2006, \aap, 452, 237

\bibitem[{{Absil} {et~al.}(2008{\natexlab{b}}){Absil}, {di Folco},
  {M{\'e}rand}, {Augereau}, {Coud{\'e} Du Foresto}, {Defr{\`e}re}, {Kervella},
  {Aufdenberg}, {Desort}, {Ehrenreich}, {Lagrange}, {Montagnier}, {Olofsson},
  {Ten Brummelaar}, {McAlister}, {Sturmann}, {Sturmann}, \&
  {Turner}}]{Absil:2007b}
{Absil}, O., {di Folco}, E., {M{\'e}rand}, A., {et~al.} 2008{\natexlab{b}},
  \aap, 487, 1041

\bibitem[{{Absil} {et~al.}(2009){Absil}, {Mennesson}, {Le Bouquin}, {Di Folco},
  {Kervella}, \& {Augereau}}]{Absil:2009}
{Absil}, O., {Mennesson}, B., {Le Bouquin}, J.-B., {et~al.} 2009, ArXiv
  e-prints

\bibitem[{{Akeson} {et~al.}(2009){Akeson}, {Ciardi}, {Millan-Gabet}, {Merand},
  {Folco}, {Monnier}, {Beichman}, {Absil}, {Aufdenberg}, {McAlister},
  {Brummelaar}, {Sturmann}, {Sturmann}, \& {Turner}}]{Akeson:2009}
{Akeson}, R.~L., {Ciardi}, D.~R., {Millan-Gabet}, R., {et~al.} 2009, \apj, 691,
  1896

\bibitem[{{Alonso} {et~al.}(2008){Alonso}, {Auvergne}, {Baglin}, {Ollivier},
  {Moutou}, {Rouan}, {Deeg}, {Aigrain}, {Almenara}, {Barbieri}, {Barge},
  {Benz}, {Bord{\'e}}, {Bouchy}, {De la Reza}, {Deleuil}, {Dvorak}, {Erikson},
  {Fridlund}, {Gillon}, {Gondoin}, {Guillot}, {Hatzes}, {H{\'e}brard},
  {Kabath}, {Jorda}, {Lammer}, {L{\'e}ger}, {Llebaria}, {Loeillet}, {Magain},
  {Mayor}, {Mazeh}, {P{\"a}tzold}, {Pepe}, {Pont}, {Queloz}, {Rauer},
  {Shporer}, {Schneider}, {Stecklum}, {Udry}, \& {Wuchterl}}]{Alonso:2008}
{Alonso}, R., {Auvergne}, M., {Baglin}, A., {et~al.} 2008, \aap, 482

\bibitem[{{Angel} \& {Woolf}(1997)}]{Angel:1997}
{Angel}, J.~R.~P. \& {Woolf}, N.~J. 1997, \apj, 475, 373

\bibitem[{{Barge} {et~al.}(2008){Barge}, {Baglin}, {Auvergne}, {Rauer},
  {Leger}, {Schneider}, {Pont}, {Aigrain}, {Almenara}, {Alonso}, {Barbieri},
  {Borde}, {Bouchy}, {Deeg}, {De la Reza}, {Deleuil}, {Dvorak}, {Erikson},
  {Fridlund}, {Gillon}, {Gondoin}, {Guillot}, {Hatzes}, {Hebrard}, {Jorda},
  {Kabath}, {Lammer}, {Llebaria}, {Loeillet}, {Magain}, {Mazeh}, {Moutou},
  {Ollivier}, {Patzold}, {Queloz}, {Rouan}, {Shporer}, \&
  {Wuchterl}}]{Barge:2008}
{Barge}, P., {Baglin}, A., {Auvergne}, M., {et~al.} 2008, \aap, 482

\bibitem[{{Beichman} {et~al.}(2006){Beichman}, {Bryden}, {Stapelfeldt},
  {Gautier}, {Grogan}, {Shao}, {Velusamy}, {Lawler}, {Blaylock}, {Rieke},
  {Lunine}, {Fischer}, {Marcy}, {Greaves}, {Wyatt}, {Holland}, \&
  {Dent}}]{Beichman:2006b}
{Beichman}, C.~A., {Bryden}, G., {Stapelfeldt}, K.~R., {et~al.} 2006, \apj,
  652, 1674

\bibitem[{{Borucki} {et~al.}(2007){Borucki}, {Koch}, {Lissauer}, {Basri},
  {Brown}, {Caldwell}, {Jenkins}, {Caldwell}, {Christensen-Dalsgaard},
  {Cochran}, {Dunham}, {Gautier}, {Geary}, {Latham}, {Sasselov}, {Gilliland},
  {Howell}, {Monet}, \& {Batalha}}]{Borucki:2007}
{Borucki}, W.~J., {Koch}, D.~G., {Lissauer}, J., {et~al.} 2007, in Astronomical
  Society of the Pacific Conference Series, Vol. 366, Transiting Extrapolar
  Planets Workshop, ed. C.~{Afonso}, D.~{Weldrake}, \& T.~{Henning}, 309--+

\bibitem[{{Bracewell}(1978)}]{Bracewell:1978}
{Bracewell}, R.~N. 1978, \nat, 274, 780

\bibitem[{{Carle}(2005)}]{Carle:2005}
{Carle}, E. 2005, {EMMA configuration: evaluation of optical performances},
  Tech. Rep. Issue 1, ESA (SCI-A/279)

\bibitem[{{Chazelas} {et~al.}(2006){Chazelas}, {Brachet}, {Bord{\'e}},
  {Mennesson}, {Ollivier}, {Absil}, {Lab{\`e}que}, {Valette}, \&
  {L{\'e}ger}}]{Chazelas:2006}
{Chazelas}, B., {Brachet}, F., {Bord{\'e}}, P., {et~al.} 2006, \ao, 45, 984

\bibitem[{{d'Arcio}(2005)}]{Darcio:2005}
{d'Arcio}, L. 2005

\bibitem[{{Defr{\`e}re} {et~al.}(2008{\natexlab{a}}){Defr{\`e}re}, {Absil},
  {Coud{\'e} Du Foresto}, {Danchi}, \& {den Hartog}}]{Defrere:2007c}
{Defr{\`e}re}, D., {Absil}, O., {Coud{\'e} Du Foresto}, V., {Danchi}, W.~C., \&
  {den Hartog}, R. 2008{\natexlab{a}}, \aap, 490, 435

\bibitem[{{Defr{\`e}re} {et~al.}(2008{\natexlab{b}}){Defr{\`e}re}, {Lay}, {den
  Hartog}, \& {Absil}}]{Defrere:2008}
{Defr{\`e}re}, D., {Lay}, O., {den Hartog}, R., \& {Absil}, O.
  2008{\natexlab{b}}, in Proc. SPIE, Vol. 7013

\bibitem[{{den Hartog}(2005{\natexlab{a}})}]{Den_Hartog:2005b}
{den Hartog}, R. 2005{\natexlab{a}}, {DARWIN science performance prediction},
  Tech. Rep. Issue 1, ESA (SCI-A/300)

\bibitem[{{den Hartog}(2005{\natexlab{b}})}]{Den_Hartog:2005}
{den Hartog}, R. 2005{\natexlab{b}}, {The DARWINsim science simulator}, Tech.
  Rep. Issue 1, ESA (SCI-A/297)

\bibitem[{{Dermott} {et~al.}(1994){Dermott}, {Jayaraman}, {Xu}, {Gustafson}, \&
  {Liou}}]{Dermott:1994}
{Dermott}, S.~F., {Jayaraman}, S., {Xu}, Y.~L., {Gustafson}, B.~A.~S., \&
  {Liou}, J.~C. 1994, in , 719--723

\bibitem[{{Dermott} {et~al.}(1985){Dermott}, {Nicholson}, {Burns}, \&
  {Houck}}]{Dermott:1985}
{Dermott}, S.~F., {Nicholson}, P.~D., {Burns}, J.~A., \& {Houck}, J.~R. 1985,
  in Astrophysics and Space Science Library, Vol. 119, IAU Colloq. 85:
  Properties and Interactions of Interplanetary Dust, ed. R.~H. {Giese} \&
  P.~{Lamy}, 395--409

\bibitem[{{Di Folco} {et~al.}(2007){Di Folco}, {Absil}, {Augereau},
  {M{\'e}rand}, {Coud{\'e} Du Foresto}, {Th{\'e}venin}, {Defr{\`e}re},
  {Kervella}, {Ten Brummelaar}, {McAlister}, {Ridgway}, {Sturmann}, {Sturmann},
  \& {Turner}}]{Difolco:2007}
{Di Folco}, E., {Absil}, O., {Augereau}, J.-C., {et~al.} 2007, \aap, 475, 243

\bibitem[{{Dohnanyi}(1969)}]{Dohnanyi:1969}
{Dohnanyi}, J.~S. 1969, \jgr, 74, 2531

\bibitem[{{Fridlund} {et~al.}(2006){Fridlund}, {d'Arcio}, {den Hartog}, \&
  {Karlsson}}]{Fridlund:2006}
{Fridlund}, C.~V.~M., {d'Arcio}, L., {den Hartog}, R., \& {Karlsson}, A. 2006,
  in Proc. SPIE, Vol. 6268

\bibitem[{{Fridlund}(2005)}]{Fridlund:2005}
{Fridlund}, M. 2005, {Darwin Science Requirements Document}, Tech. Rep. Issue
  5, ESA (SCI-A)

\bibitem[{{Gabor} {et~al.}(2008){Gabor}, {Chazelas}, {Brachet}, {Ollivier},
  {Decaudin}, {Jacquinod}, {Lab{\`e}que}, \& {L{\'e}ger}}]{Gabor:2008}
{Gabor}, P., {Chazelas}, B., {Brachet}, F., {et~al.} 2008, \aap, 483

\bibitem[{{Gillon} {et~al.}(2007){Gillon}, {Pont}, {Demory}, {Mallmann},
  {Mayor}, {Mazeh}, {Queloz}, {Shporer}, {Udry}, \& {Vuissoz}}]{Gillon:2007}
{Gillon}, M., {Pont}, F., {Demory}, B.-O., {et~al.} 2007, \aap, 472, L13

\bibitem[{{Greaves} {et~al.}(2005){Greaves}, {Holland}, {Wyatt}, {Dent},
  {Robson}, {Coulson}, {Jenness}, {Moriarty-Schieven}, {Davis}, {Butner},
  {Gear}, {Dominik}, \& {Walker}}]{Greaves:2005}
{Greaves}, J.~S., {Holland}, W.~S., {Wyatt}, M.~C., {et~al.} 2005, \apjl, 619,
  L187

\bibitem[{{Hinkley} {et~al.}(2007){Hinkley}, {Oppenheimer}, {Soummer},
  {Sivaramakrishnan}, {Roberts}, {Kuhn}, {Makidon}, {Perrin}, {Lloyd},
  {Kratter}, \& {Brenner}}]{Hinkley:2007}
{Hinkley}, S., {Oppenheimer}, B.~R., {Soummer}, R., {et~al.} 2007, \apj, 654,
  633

\bibitem[{{Hinz} {et~al.}(2008){Hinz}, {Bippert-Plymate}, {Breuninger},
  {Connors}, {Duffy}, {Esposito}, {Hoffmann}, {Kim}, {Kraus}, {McMahon},
  {Montoya}, {Nash}, {Durney}, {Solheid}, {Tozzi}, \&
  {Vaitheeswaran}}]{Hinz:2008b}
{Hinz}, P.~M., {Bippert-Plymate}, T., {Breuninger}, A., {et~al.} 2008, in Proc.
  SPIE, Vol. 7013

\bibitem[{{Kalas} {et~al.}(2005){Kalas}, {Graham}, \& {Clampin}}]{Kalas:2005}
{Kalas}, P., {Graham}, J.~R., \& {Clampin}, M. 2005, in , 1067--1070

\bibitem[{{Kaltenegger} {et~al.}(2008){Kaltenegger}, {Eiroa}, \&
  {Fridlund}}]{Kaltenegger:2008}
{Kaltenegger}, L., {Eiroa}, C., \& {Fridlund}, C.~V.~M. 2008, ArXiv e-prints

\bibitem[{{Karlsson} {et~al.}(2004){Karlsson}, {Wallner}, {Perdigues Armengol},
  \& {Absil}}]{Karlsson:2004}
{Karlsson}, A.~L., {Wallner}, O., {Perdigues Armengol}, J.~M., \& {Absil}, O.
  2004, in Proc. SPIE, ed. W.~A. {Traub}, Vol. 5491, 831--+

\bibitem[{{Kelsall} {et~al.}(1998){Kelsall}, {Weiland}, {Franz}, {Reach},
  {Arendt}, {Dwek}, {Freudenreich}, {Hauser}, {Moseley}, {Odegard},
  {Silverberg}, \& {Wright}}]{Kelsall:1998}
{Kelsall}, T., {Weiland}, J.~L., {Franz}, B.~A., {et~al.} 1998, \apj, 508, 44

\bibitem[{{Ksendzov} {et~al.}(2007){Ksendzov}, {Lay}, {Martin}, {Sanghera},
  {Busse}, {Kim}, {Pureza}, {Nguyen}, \& {Aggarwal}}]{Ksendzov:2007}
{Ksendzov}, A., {Lay}, O., {Martin}, S., {et~al.} 2007, \ao, 46, 7957

\bibitem[{{Ksendzov} {et~al.}(2008){Ksendzov}, {Lay}, {Martin}, {Sanghera},
  {Busse}, {Kim}, {Pureza}, {Nguyen}, \& {Aggarwal}}]{Ksendzov:2008}
{Ksendzov}, A., {Lay}, O., {Martin}, S., {et~al.} 2008, \ao, 46, 7957

\bibitem[{{Kuchner} \& {Holman}(2003)}]{Kuchner:2003}
{Kuchner}, M.~J. \& {Holman}, M.~J. 2003, \apj, 588, 1110

\bibitem[{{Landgraf} \& {Jehn}(2001)}]{Landgraf:2001}
{Landgraf}, M. \& {Jehn}, R. 2001, \apss, 278, 357

\bibitem[{{Lane} {et~al.}(2006){Lane}, {Muterspaugh}, \& {Shao}}]{Lane:2006}
{Lane}, B.~F., {Muterspaugh}, M.~W., \& {Shao}, M. 2006, \apj, 648, 1276

\bibitem[{{Lawson} {et~al.}(2008){Lawson}, {Lay}, {Martin}, {Peters},
  {Gappinger}, {Ksendzov}, {Scharf}, {Booth}, {Beichman}, {Serabyn},
  {Johnston}, \& {Danchi}}]{Lawson:2008}
{Lawson}, P.~R., {Lay}, O.~P., {Martin}, S.~R., {et~al.} 2008, in Proc. SPIE,
  Vol. 7013

\bibitem[{{Lay}(2004)}]{Lay:2004}
{Lay}, O.~P. 2004, \ao, 43, 6100

\bibitem[{{Lay}(2005)}]{Lay:2005}
{Lay}, O.~P. 2005, \ao, 44, 5859

\bibitem[{{Lay}(2006)}]{Lay:2006}
{Lay}, O.~P. 2006, in Proc. SPIE, Vol. 6268

\bibitem[{{Lay} {et~al.}(2007){Lay}, {Martin}, \& L.}]{Lay:2007}
{Lay}, O.~P., {Martin}, S.~R., \& L., H.~S. 2007, in Proc. SPIE, Vol. 6693

\bibitem[{{L\'eger} \& {Herbst}(2007)}]{Leger:2007}
{L\'eger}, A. \& {Herbst}, T. 2007, ArXiv e-prints, 707

\bibitem[{{L\'eger} {et~al.}(subm.){L\'eger}, {Rouan}, {Schneider}, {Alonso},
  \& {Samuel}}]{Leger:2009}
{L\'eger}, A., {Rouan}, D., {Schneider}, J., {Alonso}, B., \& {Samuel}, E.
  subm., \aap

\bibitem[{{Liou} \& {Zook}(1999)}]{Liou:1999}
{Liou}, J.-C. \& {Zook}, H.~A. 1999, \aj, 118, 580

\bibitem[{{Macintosh} {et~al.}(2003){Macintosh}, {Becklin}, {Kaisler},
  {Konopacky}, \& {Zuckerman}}]{Macintosh:2003}
{Macintosh}, B.~A., {Becklin}, E.~E., {Kaisler}, D., {Konopacky}, Q., \&
  {Zuckerman}, B. 2003, \apj, 594, 538

\bibitem[{{Mawet} {et~al.}(2007){Mawet}, {Hanot}, {Lenaers}, {Riaud},
  {Defr{\`e}re}, {Vandormael}, {Loicq}, {Fleury}, {Plesseria}, {Surdej}, \&
  {Habraken}}]{Mawet:2007}
{Mawet}, D., {Hanot}, C., {Lenaers}, C., {et~al.} 2007, Optics Express, 15,
  12850

\bibitem[{{Mayor} {et~al.}(2009){Mayor}, {Bonfils}, {Forveille}, {Delfosse},
  {Udry}, {Bertaux}, {Beust}, {Bouchy}, {Lovis}, {Pepe}, {Perrier}, {Queloz},
  \& {Santos}}]{Mayor:2009b}
{Mayor}, M., {Bonfils}, X., {Forveille}, T., {et~al.} 2009, \aap, in press

\bibitem[{{Mayor} \& {Queloz}(1995)}]{Mayor:1995}
{Mayor}, M. \& {Queloz}, D. 1995, \nat, 378, 355

\bibitem[{{Mennesson} \& {Mariotti}(1997)}]{Mennesson:1997}
{Mennesson}, B. \& {Mariotti}, J.~M. 1997, Icarus, 128, 202

\bibitem[{{Moro-Mart{\'{\i}}n} \& {Malhotra}(2005)}]{Moro-Martin:2005}
{Moro-Mart{\'{\i}}n}, A. \& {Malhotra}, R. 2005, \apj, 633, 1150

\bibitem[{{Moutou} {et~al.}(2005){Moutou}, {Pont}, {Barge}, {Aigrain},
  {Auvergne}, {Blouin}, {Cautain}, {Erikson}, {Guis}, {Guterman}, {Irwin},
  {Lanza}, {Queloz}, {Rauer}, {Voss}, \& {Zucker}}]{Moutou:2005}
{Moutou}, C., {Pont}, F., {Barge}, P., {et~al.} 2005, \aap, 437, 355

\bibitem[{{Ozernoy} {et~al.}(2000){Ozernoy}, {Gorkavyi}, {Mather}, \&
  {Taidakova}}]{Ozernoy:2000}
{Ozernoy}, L.~M., {Gorkavyi}, N.~N., {Mather}, J.~C., \& {Taidakova}, T.~A.
  2000, \apjl, 537, L147

\bibitem[{{Peters.} {et~al.}(2009){Peters.}, {Gappinger}, {Lawson}, \&
  {Lay}}]{Peters:2009}
{Peters.}, R., {Gappinger}, R., {Lawson}, P., \& {Lay}, O. 2009, JPL, Document
  D-60326

\bibitem[{{Reach} {et~al.}(1995){Reach}, {Franz}, {Weiland}, {Hauser},
  {Kelsall}, {Wright}, {Rawley}, {Stemwedel}, \& {Spiesman}}]{Reach:1995}
{Reach}, W.~T., {Franz}, B.~A., {Weiland}, J.~L., {et~al.} 1995, \nat, 374, 521

\bibitem[{{Reche} {et~al.}(2008){Reche}, {Beust}, {Augereau}, \&
  {Absil}}]{Reche:2008}
{Reche}, R., {Beust}, H., {Augereau}, J.-C., \& {Absil}, O. 2008, \aap, 480,
  551

\bibitem[{{Richardson} {et~al.}(2007){Richardson}, {Deming}, {Horning},
  {Seager}, \& {Harrington}}]{Richardson:2007}
{Richardson}, L.~J., {Deming}, D., {Horning}, K., {Seager}, S., \&
  {Harrington}, J. 2007, \nat, 445, 892

\bibitem[{{Robertson}(1937)}]{Robertson:1937}
{Robertson}, H.~P. 1937, \mnras, 97, 423

\bibitem[{{Roques} {et~al.}(1994){Roques}, {Scholl}, {Sicardy}, \&
  {Smith}}]{Roques:1994}
{Roques}, F., {Scholl}, H., {Sicardy}, B., \& {Smith}, B.~A. 1994, Icarus, 108,
  37

\bibitem[{{Ruilier} \& {Cassaing}(2001)}]{Ruilier:2001}
{Ruilier}, C. \& {Cassaing}, F. 2001, Journal of the Optical Society of America
  A, 18, 143

\bibitem[{{Schneider} {et~al.}(2009){Schneider}, {Weinberger}, {Becklin},
  {Debes}, \& {Smith}}]{Schneider:2009}
{Schneider}, G., {Weinberger}, A.~J., {Becklin}, E.~E., {Debes}, J.~H., \&
  {Smith}, B.~A. 2009, \aj, 137, 53

\bibitem[{{Serabyn}(2009)}]{Serabyn:2009}
{Serabyn}, E. 2009, \apj, 697, 1334

\bibitem[{{Serabyn} \& {Colavita}(2001)}]{Serabyn:2001}
{Serabyn}, E. \& {Colavita}, M.~M. 2001, \ao, 40, 1668

\bibitem[{{Stark} \& {Kuchner}(2008)}]{Stark:2008}
{Stark}, C.~C. \& {Kuchner}, M.~J. 2008, \apj, 686, 637

\bibitem[{{Stark} \& {Kuchner}(2009)}]{Stark:2009}
{Stark}, C.~C. \& {Kuchner}, M.~J. 2009, ArXiv e-prints

\bibitem[{{Stark} {et~al.}(2009){Stark}, {Kuchner}, {Traub}, {Monnier},
  {Serabyn}, \& {Colavita}}]{Stark:2009b}
{Stark}, C.~C., {Kuchner}, M.~J., {Traub}, W.~A., {et~al.} 2009, 41, 501

\bibitem[{{Thiebaut} \& {Mugnier}(2006)}]{Thiebaut:2006}
{Thiebaut}, E. \& {Mugnier}, L. 2006, in IAU Colloq. 200: Direct Imaging of
  Exoplanets: Science and Techniques, 547--552

\bibitem[{{Wilner} {et~al.}(2002){Wilner}, {Holman}, {Kuchner}, \&
  {Ho}}]{Wilner:2002}
{Wilner}, D.~J., {Holman}, M.~J., {Kuchner}, M.~J., \& {Ho}, P.~T.~P. 2002,
  \apjl, 569, L115

\end{thebibliography}
\bibliographystyle{aa}

\end{document}